# Chondrule heritage and thermal histories from trace element and oxygen isotope analyses of chondrules and amoeboid olivine aggregates

Emmanuel Jacquet[1], Yves Marrocchi[2]

[1]Institut de Minéralogie, de Physique des Matériaux et de Cosmochimie, CNRS & Muséum National d'Histoire Naturelle, UMR 7590, 57 rue Cuvier, 75005 Paris, France.

[2]Centre de Recherches Pétrographiques et Géochimiques, CNRS, Université de Lorraine, UMR 7358, 54501 Vandoeuvre-lès-Nancy, France.

E-mail: emmanuel.jacquet@mnhn.fr

*Abstract*

We report combined oxygen isotope and mineral-scale trace element analyses of amoeboid olivine aggregates (AOA) and chondrules in ungrouped carbonaceous chondrite Northwest Africa 5958. The trace element geochemistry of olivine in AOA, for the first time measured by LA-ICP-MS, is consistent with a condensation origin although the shallow slope of its rare earth element (REE) pattern has yet to be physically explained. Ferromagnesian silicates in type I chondrules resemble those in other carbonaceous chondrites both geochemically and isotopically, and we find a correlation between $^{16}$O enrichment and many incompatible elements in olivine. The variation in incompatible element concentrations may relate to varying amounts of olivine crystallization during a sub-isothermal stage of chondrule-forming events, the duration of which may be anticorrelated with the local solid/gas ratio if this was the determinant of oxygen isotopic ratios as recently proposed. While aqueous alteration has depleted many chondrule mesostases in REE, some chondrules show recognizable subdued group II-like patterns supporting the idea that the immediate precursors of chondrules were nebular condensates.

# Introduction

Chondrites, while recognized for decades as our chief witnesses of the early solar system, still remain poorly rationalized collections of sub-millimeter-sized inclusions: *Refractory inclusions*, comprising Calcium-Aluminum-rich Inclusions (CAI; MacPherson 2014) and the more common Amoeboid Olivine Aggregates (AOA; Krot et al. 2004), the oldest objects dated in meteorites; *chondrules*, silicate spherules apparently produced by transient high-temperature events but in an obstinately elusive context (Connolly and Desch 2004).

Half a century ago, the condensation sequence (e.g. Suess (1949); Larimer and Anders (1970); Grossman (1972)) offered the prospect of a unified model for all these objects. Assuming an originally hot (~2000 K) solar nebula, thermodynamic calculations predicted the successive appearance of the parageneses of refractory inclusions, chondrules and chondrite fine-grained matrices. While refractory inclusions, with an $^{16}$O-rich isotopic composition similar to the Sun's (McKeegan et al. 2011), indeed remain overall most consistent with initial formation as condensates, or perhaps evaporation residues—notwithstanding possible reheating events—, chondrules are hardly thought of as direct nebular condensates (as entertained e.g. by Wood (1963)) anymore since the low total pressures in the disk would make liquid condensates unstable and the cooling timescales of the inner disk would be too long to satisfy textural observations. Also, recent evidence such as FeO content, retention of volatile elements or the high frequency of compound chondrules points to strong differences between chondrule-forming environments and canonical pictures of the average protoplanetary disk (e.g. Alexander et al. (2008); Grossman et al. (2012); Fedkin and Grossman (2013); Marrocchi and Libourel (2013)). Not to mention that chondrite matrices may have been affected more by parent body alteration than by nebular processes (McSween (1979); Brearley (2014)). Clearly, condensation, while obviously an important part of the story, was not everything.

Yet all ties between chondrules and nebular condensates may not be broken. Compositionally anomalous crystals in chondrules have long been interpreted as relicts from solid precursors (Nagahara (1981); Rambaldi (1981)). While a planetary origin has been advocated for some olivine grains (Libourel and Krot (2007); Libourel and Chaussidon (2011)), a kinship of these precursors with refractory inclusions and other condensates (Russell et al. 2005) is suggested by compound

CAI/chondrule objects (Christophe Michel-Lévy (1986); Aléon and Bourot-Denise (2008); Krot et al. (2009)), (relatively $^{16}$O-rich) aluminum-rich chondrules (Russell et al. (2000); Krot and Keil (2002)), and volatility-fractionated rare earth element patterns in bulk Allende chondrules (Misawa and Nakamura 1988). More specifically, a derivation of chondrules from the chemically most similar AOAs has been found possible (Krot et al. (2004); Whattam et al. (2010)), if open-system behavior, in particular for Si, is allowed for, such as in fact suggested for type I chondrules on petrographic grounds (Hezel et al. (2003); Libourel et al. (2006)); and a relict AOA has even been reported in a type II chondrule in CO3.0 chondrite Yamato 81020 (Yurimoto and Wasson 2002).

To gain further insight, oxygen isotope and trace element analyses at the mineral scale of both chondrules and AOAs would offer a helpful comparison. Indeed, on the one hand, oxygen isotopes have long been used as an effective "map" of meteoritics (Clayton 2003), and possible refractory inclusion inheritance may still be signaled by $^{16}$O enrichment in chondrules. Trace element concentrations, on the other hand, are a sensitive marker of partitioning, whether igneous or with a gas phase, and thence formation or resetting conditions of these objects. Of course, none of this is entirely new individually. O isotope measurements are notoriously popular for chondrule minerals, in particular olivine, in recent literature (Ruzicka et al. (2007); Kita et al. (2010); Libourel and Chaussidon (2011); Rudraswami et al. (2011); Ushikubo et al. (2012); Tenner et al. (2013); Schrader et al. (2013); Tenner et al. (2015); Metzler and Pack (2016); Tenner et al. (2017)) and have been quite systematically measured in AOAs as well (see references in Krot et al. 2004)). Trace element composition of chondrule phases have been reported using SIMS in ordinary chondrites (e.g. Alexander (1994); Jones and Layne (1997); Ruzicka et al. (2008)) and lately using laser ablation inductively coupled mass spectrometry (LA-ICP-MS) in CV, CR, LL and EH chondrites (Jacquet et al. (2012); Jacquet et al. (2013); Jacquet et al. (2015b); Jacquet et al. (2015a)). As to AOAs, and specifically those of CV chondrites, while bulk compositions were reported by Grossman et al. (1979) and Russell et al. (2003), Ruzicka et al. (2012a) presented SIMS analyses, some (narrow-beam) at the mineral scale.

Any investigation of links between refractory inclusions and chondrules in a given chondrite is potentially complicated by redistributions of solids which may have occurred over the millions of years elapsed between original nebular condensation and incorporation in the parent body. This problem may however be minimized in the least fractionated (carbonaceous) chondrites whose reservoirs may have behaved as statistically closed systems (Jacquet et al. 2012), such as CI and CM chondrites. Northwest Africa (NWA) 5958 is such a meteorite. A 286 g fresh find in Morocco in 2009, it was first classified as a C3.0 UNG (Bunch et al. (2011); Ash et al. (2011)) although the multi-technique description presented by Jacquet et al. (2016) has resulted in reclassification to a weakly altered C2 ungrouped. At any rate, it is one of the least altered CM-like meteorites known. Its quite uniquely compact-textured AOAs (Jacquet et al. 2016) constitute our best chance, among all chondrites, to obtain reliable trace element analyses of AOA olivine, with LA-ICP-MS ensuring low detection limits in shorter analysis durations compared to SIMS. As to chondrules, NWA 5958 also provides an opportunity (if less specific) to extend the mineral trace element dataset in the literature to the CM-CO clan (*sensu lato*), the most abundant clan among carbonaceous chondrites. It is thus in this meteorite that we have elected to carry out combined trace element and O isotope analyses (by LA-ICP-MS and SIMS, respectively) in chondrules and AOAs, which we report and discuss in this paper.

## Samples and methods

We studied three thick sections of Northwest Africa 5958 from the meteorite collection of the Muséum National d'Histoire Naturelle de Paris (MNHN), specifically NWA 5958-1, NWA 5958-5 and

NWA 5958-6 (object names, such as "N5-4", have prefixes "N1", "N5" or "N6" depending on their host section). 26 chondrules and 8 refractory inclusions were selected for analysis (Table EA-1). In addition, a large AOA, called M1, from CO3.2 chondrite Miller Range (MIL) 07342—specifically thick section MIL 07342,9 from NASA's Antarctic Meteorite Collection—, similar in texture to NWA 5958 AOAs, was also studied for petrography and chemistry.

The objects of interest were examined in optical and scanning electron microscopy (SEM—a JEOL JSM-840A with EDAX EDS detector and a JEOL JSM6610-Lv with an Oxford/SDD EDS detector). X-ray maps allowed apparent mineral modes of chondrules to be calculated using the JMicrovision software (www.jmicrovision.com). Minor and major element concentrations of documented chondrules were obtained with a Cameca SX-100 electron microprobe (EMP) at the Centre de Microanalyse de Paris VI (CAMPARIS).

Trace element analyses of selected objects (chondrules and refractory inclusions) were performed by LA-ICP-MS at the University of Montpellier II. The laser ablation system was a GeoLas $Q^+$ platform with an Excimer CompEx 102 laser and was coupled to a ThermoFinnigan Element XR mass spectrometer. The ICP-MS was operated at 1350 W and tuned daily to produce maximum sensitivity for the medium and high masses, while keeping the oxide production rate low ($^{248}$ThO/$^{232}$Th ≤ 1%). Ablations were performed in pure He-atmosphere (0.65 ± 0.05 L/min) mixed before entering the torch with a flow of Ar (≈ 1.00 ± 0.05 L/min). Laser ablation conditions were: fluences ca. 12 J/cm² with pulse frequencies between 5 and 10 Hz were used and spot sizes of 26-102 μm, 51 μm being a typical value. With such energy fluences, depth speed for silicates is about 1 μm/s. Each analysis consisted of 4 min on background analyses (laser off) and 40 s of ablation (laser on). Data reduction was carried out using the GLITTER software (Griffin et al. 2008). Internal standard was Si (Ca for augite), known from EMP analyses. The NIST 612 glass (Pearce et al. 1997) was used as an external standard. Examination of the time-resolved GLITTER signal allowed exclusion of contaminating phases if those appeared during ablation; the absence of such phases from the start was checked by SEM imaging of ablation craters (compared with prior images) and comparison with EMP data (example craters are shown in the Electronic Annex). For each object, as in our previous studies, a geometric (rather than arithmetic) averaging was used to calculate mean concentrations in olivine and pyroxene in order to minimize the impact of possible undetected contamination by incompatible element-rich phases (still, the instrumental uncertainty already accounts for a median proportion (among all elements) of 60 % of the average intrachondrule relative standard deviation). In the following text and in the figures, and unless otherwise noted, the data reported will be *object means*, that is, for each phase, the average of the different (successful) analyses performed on that phase in a given object; those means are also the basis for the calculated overall average or standard deviation presented. A few bulk chondrules and chondrule groundmasses were subjected to bulk analyses with broad laser beams where the internal standard Si was determined from modal reconstruction and EMP analyses.

*In situ* oxygen isotopic measurements were performed at the CRPG-CNRS (Nancy, France) using a Cameca 1280HR secondary ion mass spectrometer (SIMS). We used a $Cs^+$ primary ion beam (15 keV, ~0.4nA, with ~10 μm diameter) and an electron flow parallel to the sample surface to neutralize charge excesses. $^{16}O^-$, $^{17}O^-$, and $^{18}O^-$ ions were measured in multi-collection mode with two Faraday cups (FC) for $^{16}O^-$ and $^{18}O^-$ and one Electron Multiplier (EM) for $^{17}O^-$ and. In order to remove $^{16}OH^-$ interference in the $^{17}O^-$ peak and to maximize the flatness of the $^{16}O^-$ and $^{18}O^-$ peaks, entrance and exit slits were adjusted to achieve a Mass Resolving Power of ≈ 7000 for $^{17}O^-$ on the central EM and ≈ 2500 on the off-axis FCs. As an additional safeguard against the $^{16}OH^-$ interference in the $^{17}O^-$ peak, we used a $N_2$ trap to reduce the pressure in the analysis chamber (i.e., < 3×10$^{-9}$ mbar). The EM Dead-time correction was applied to mass $^{17}O^-$. Each acquisition lasted for 420 s, consisting of 300 s of pre-sputtering and 120 s of simultaneous continuous measurement over the three collectors. We used three terrestrial standard materials (San Carlos olivine, quartz, and basaltic glass) to (i) define the terrestrial fractionation line (TFL) and (ii) to correct the matrix effect on instrumental mass fractionation (IMF) for olivine.

The isotopic compositions are expressed in standard δ-notation, relative to Vienna standard mean ocean water (VSMOW): $\delta^{18}O = (^{18}O/^{16}O)_{sample}/(^{18}O/^{16}O)_{VSMOW} - 1$ and $\delta^{17}O = (^{17}O/^{16}O)_{sample}/(^{17}O/^{16}O)_{VSMOW} - 1$ (both to be expressed in ‰). $\Delta^{17}O$ represents deviation from the TFL and is defined as $\delta^{17}O - 0.52 \times \delta^{18}O$. The 2σ errors were 0.1-2.4 ‰ for $\delta^{17}O$, 0.1-1.8 ‰ for $\delta^{18}O$, and 0.2-2.0 ‰ for $\Delta^{17}O$. The error on $\Delta^{17}O$ was calculated by propagating the errors on $\delta^{17}O$ and $\delta^{18}O$, and the standard deviation of $\Delta^{17}O$ values of the four terrestrial standards.

## Results

### Petrography

The petrography of the inclusions in NWA 5958 has already been described in Jacquet et al. (2016) and thus only a brief summary will be given here. Images of selected objects are provided in Fig. 1 and 2 (with further examples in Fig. 2 and 4 of Jacquet et al. (2016)). Petrographic data of analyzed objects are summarized in the Electronic Annex (Table EA1).

Most chondrules are of type I (that is, with Fe/(Mg+Fe) < 10 mol% in their ferromagnesian silicates), with porphyritic texture (mostly olivine-pyroxene (POP), with some PP and PO also encountered) mineralogically zoned from olivine- (and mesostasis-)dominated interiors to enstatite-rich margins. Some chondrules are merely single olivine grains surrounded by some enstatite but form a textural continuum with their *bona fide* porphyritic counterparts. Type II chondrules are generally PO-textured ; some of their mesostasis being peppered with abundant < 10 µm fayalitic olivine crystals (the association of the two we shall call "groundmass" hereafter; e.g. Fig. 2f of Jacquet et al. (2016) or Fig. 1f herein) similar to CO3.0 chondrules described by Wasson and Rubin (2003). Some barred olivine (BO) and cryptocrystalline (C) chondrules have also been analyzed. Mesostasis in virtually all chondrules has been altered to phyllosilicates intergrown with (more resilient) augite laths (only as isolated pockets in olivine crystals may glass be still preserved, one having been analyzed in type II chondrule N5-1 shown in Fig. 2f of Jacquet et al. (2016)). EMP traverses (see Table EA2; Fig. 3) show type II chondrule olivine to be normally zoned, with incompatible minor elements concentrations increasing toward the rims, with N6-7 (Fig. 1h of Jacquet et al. (2016)) exhibiting one oscillation (Fig. 3f). Type I chondrule olivine is often unzoned but large, olivine grains show enrichments in Fe and Cr along with *depletions* in Ca and Al (relatively high in the interior) near the margin (with a marked inflection in N6-16; Fig. 3d), a type of zoning noticed in Semarkona by Jones and Scott (1989) and apparent in Niger (CM) and CO3.0 analyses by Desnoyers (1980) and Scott and Jones (1990), respectively. 19 type I and 7 type II chondrules were analyzed in this study.

Amoeboid Olivine Aggregates (AOA) have mostly compact textures, with continuous olivine (which could be actually polycrystalline judging from TEM microstructural studies and BSE images of ALHA 77307 (CO3.0) AOAs by Han and Brearley (2015)), and altered Ca-Al-rich vermicular patches where diopside is usually the only preserved primary mineral. An EMP traverse across AOA N1-14 (pictured in Fig. 2a) indicates CaO concentrations up to 0.2 wt% at the core decreasing to 0.1 wt% near the (slightly Fe, Cr-enriched) rim (Table EA2). The AOA analyzed in MIL 07342, M1 (Fig. 2b), has a similar petrography but without aqueous alteration, hence the survival of anorthite. Although as such, these objects are hardly "aggregates" anymore (Chizmadia et al. 2002), we shall continue to refer to them as AOAs (and not risk confusion with the very different "agglomeratic objects" studied e.g. by Ruzicka et al. (2012b)!) given their continuum with normal AOAs (see the full range of textures e.g. in Krot et al. (2004)) presumably owing to variable degrees of post-formation sintering. Calcium-aluminum-rich inclusions (CAIs) are usually aggregations of spinel nodules rimmed by diopside, alteration products and occasional perovskite or olivine (similar to the Mighei CAIs described by MacPherson and Davis (1994)). One CAI, N6-18 (Fig. 2b), is a simple monomineralic spherule of aluminian-titanian diopside. 6 AOA and 3 CAI were analyzed in this study.

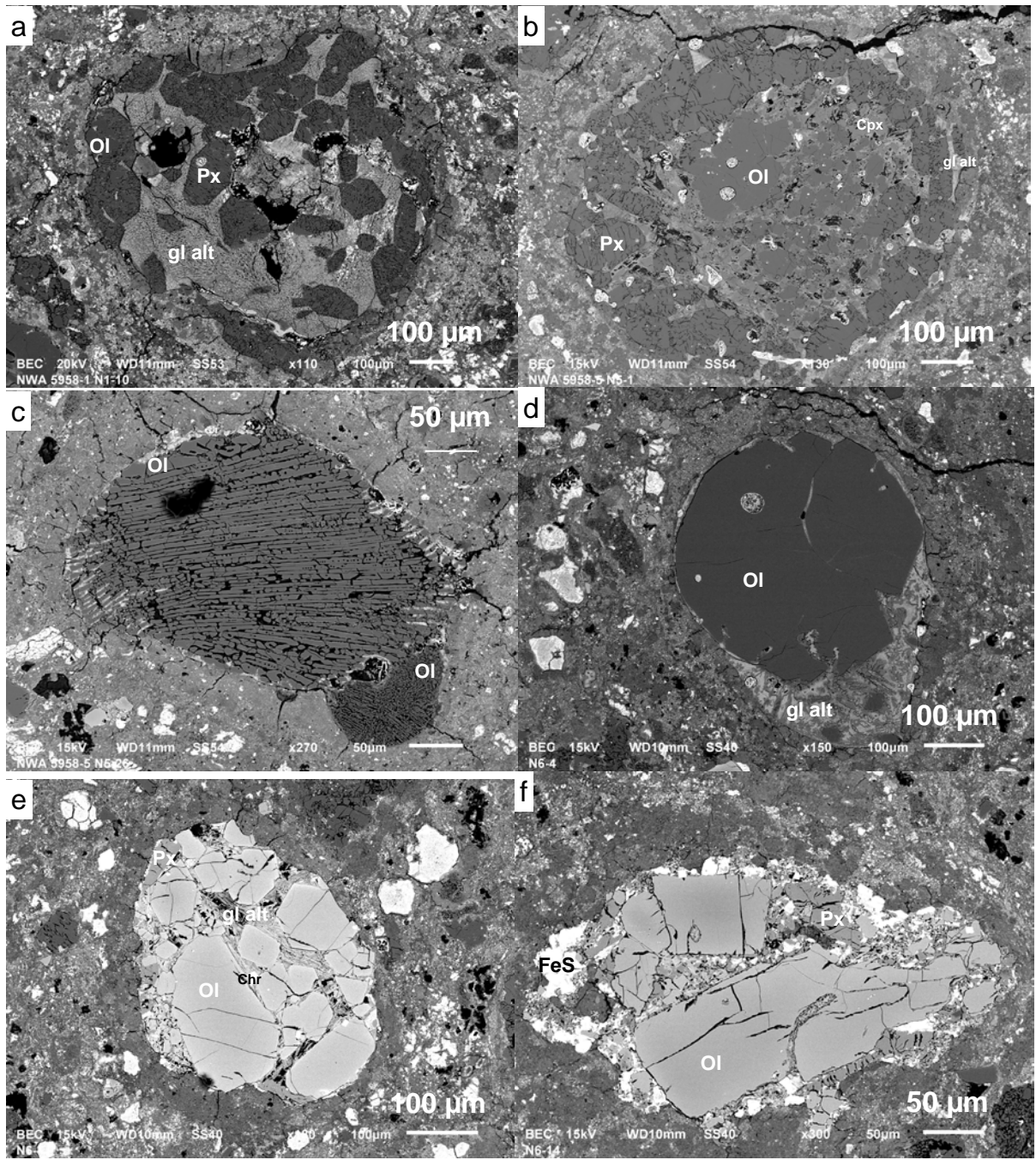

**Figure 1**: Back-scattered electron images of NWA 5958 chondrules. *Type I chondrules:* (a) PP chondrule N1-10. The euhedral enstatite crystals are embedded in an abundant mesostasis with augite laths having survived alteration. A prominent dent at the top of the chondrule is lined by a thicker-than-average portion of the chondrule's fine-grained rim. (b) POP chondrule with classical type I chondrule mineralogical zoning, with olivine crystals—often interspersed with augite-rich mesostasis, but forming a cluster near the upper left—dominating the center and enstatite crystals

near the margin and subparallel to it. (c) Compound chondrule N5-26. The main component (N5-26a) is an elongate BO chondrule fragment with little mesostasis between the forsteritic bars. The adhesion (N5-26b) on the lower right side is similarly a BO chondrule, with at least two sets of parallel bars much thinner (about a micron) than its sibling's (~5-10 μm). (d) PO chondrule N6-5, essentially a round ~0.4 mm diameter olivine grain embedded in an asymmetric droplet of altered mesostasis (some of it invading embayments or fractures within the main grain), mostly visible on the bottom side, with abundant augite, whether as laths or rims around olivine. *Type II chondrules:* (e) PO chondrule N6-4, with iron-rich euhedral olivine (Fo$_{54-59}$) set in acicular altered mesostasis along with occasional smaller pyroxene crystals with Ca-rich overgrowths. (f) POP chondrule N6-15. Two euhedral olivine crystals (with altered mesostasis inclusions or embayments) stand out in an olivine-pyroxene-altered glass groundmass, with some sulfide lining the irregular outlines of the elongated chondrule. Abbreviations: Ol = olivine, Px = low-Ca pyroxene, Cpx = Ca-rich pyroxene, Chr = chromite, gl alt = altered mesostasis, FeS = iron sulfide.

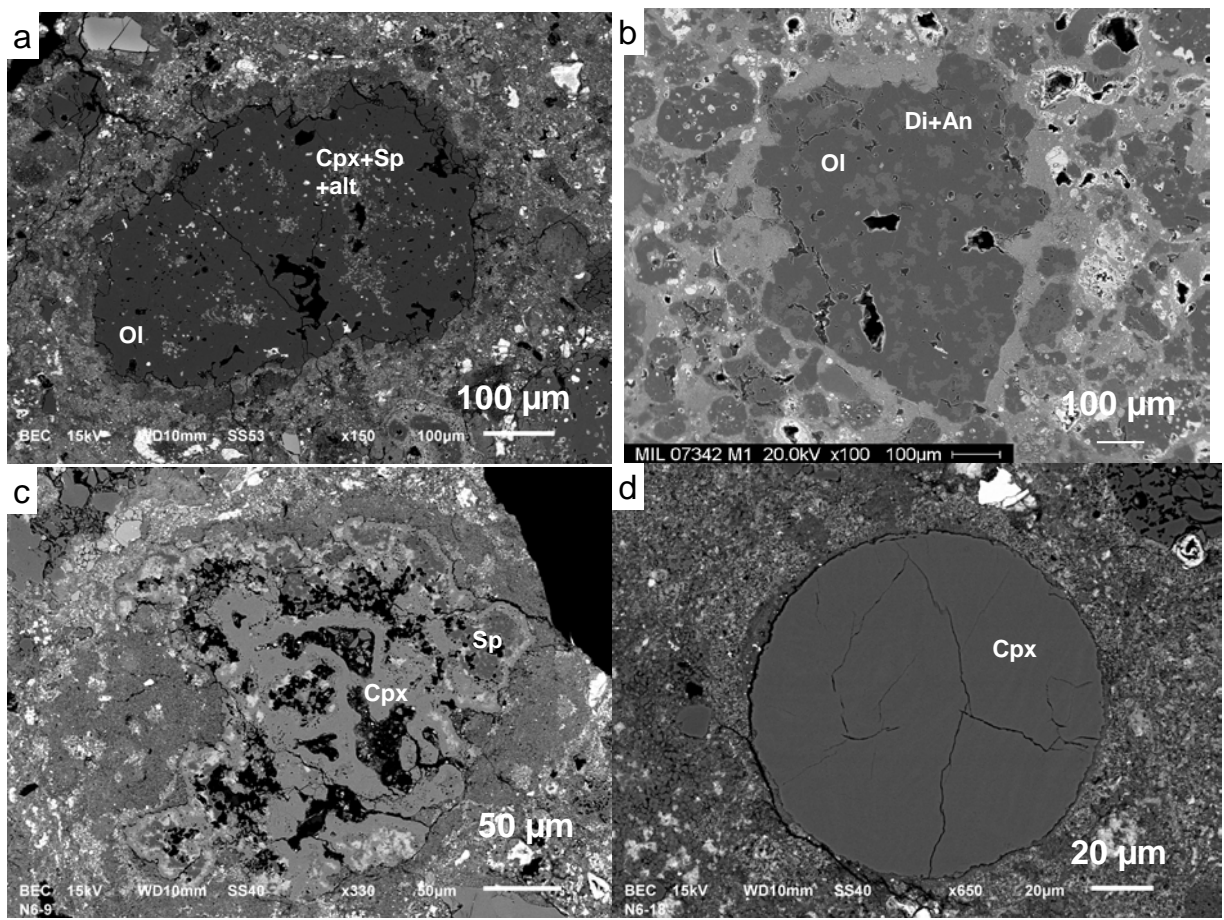

**Figure 2**: Back-scattered electron images of analyzed refractory inclusions (in NWA 5958 except (b)). (a) AOA N1-14, essentially a compact, continuous olivine body 0.5 mm in its longest dimension, with irregular patches of diopside and Al-rich alteration phases. (b) AOA M1 (in MIL 07342), similar in texture to the previous one, but with larger Ca-Al-rich patches, preserved here as intergrowths of anorthite and diopside. Occasional abrupt interruptions of the latter at the edge suggest that M1, although already 0.8 mm wide, is a fragment of an even larger body. (c) Calcium-aluminum-rich inclusion N6-9, with spinel clusters surrounded by 10-20 μm thick diopside rims. Alteration phases are whitish in color whereas the dark holes result from plucking during section preparation. (d) ~120 μm diameter Al-Ti-rich diopside spherule N6-18, homogeneous in appearance and mineral chemistry. Abbreviations as previously with Sp = spinel, An = anorthite.

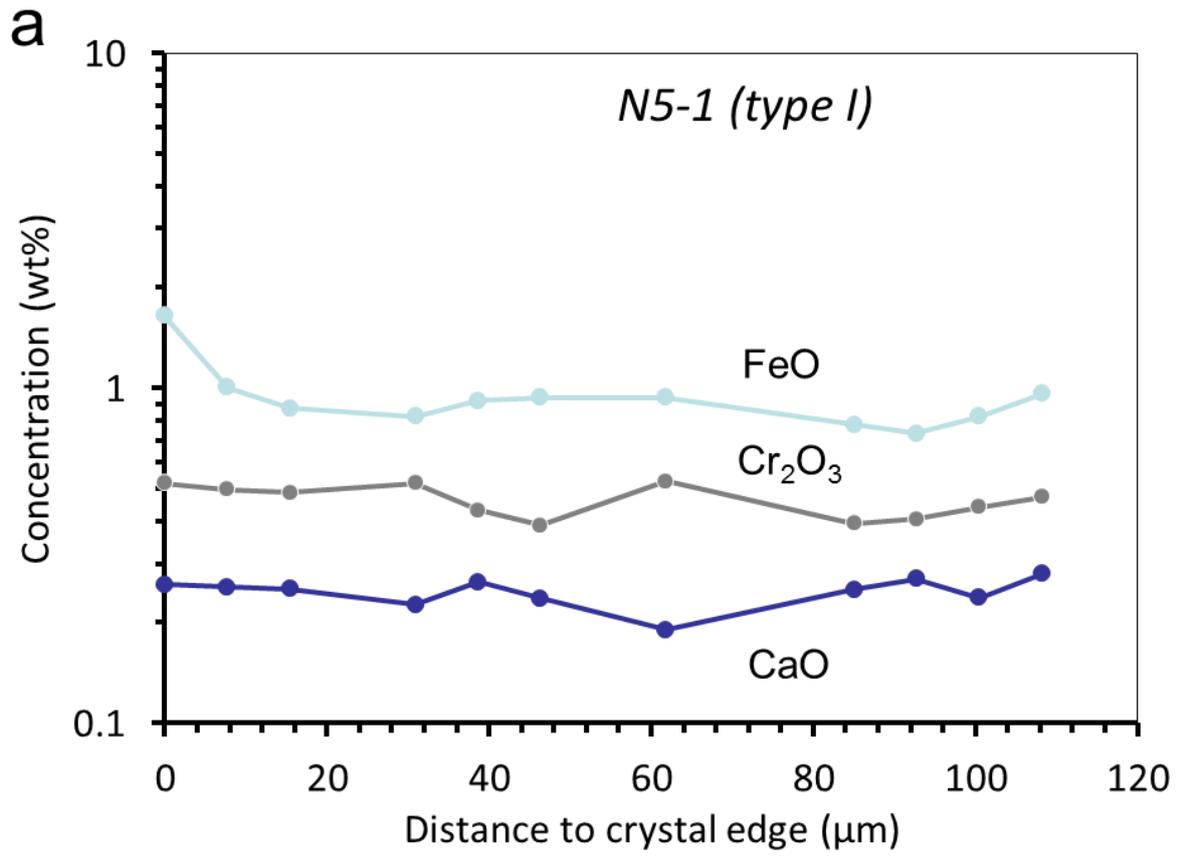
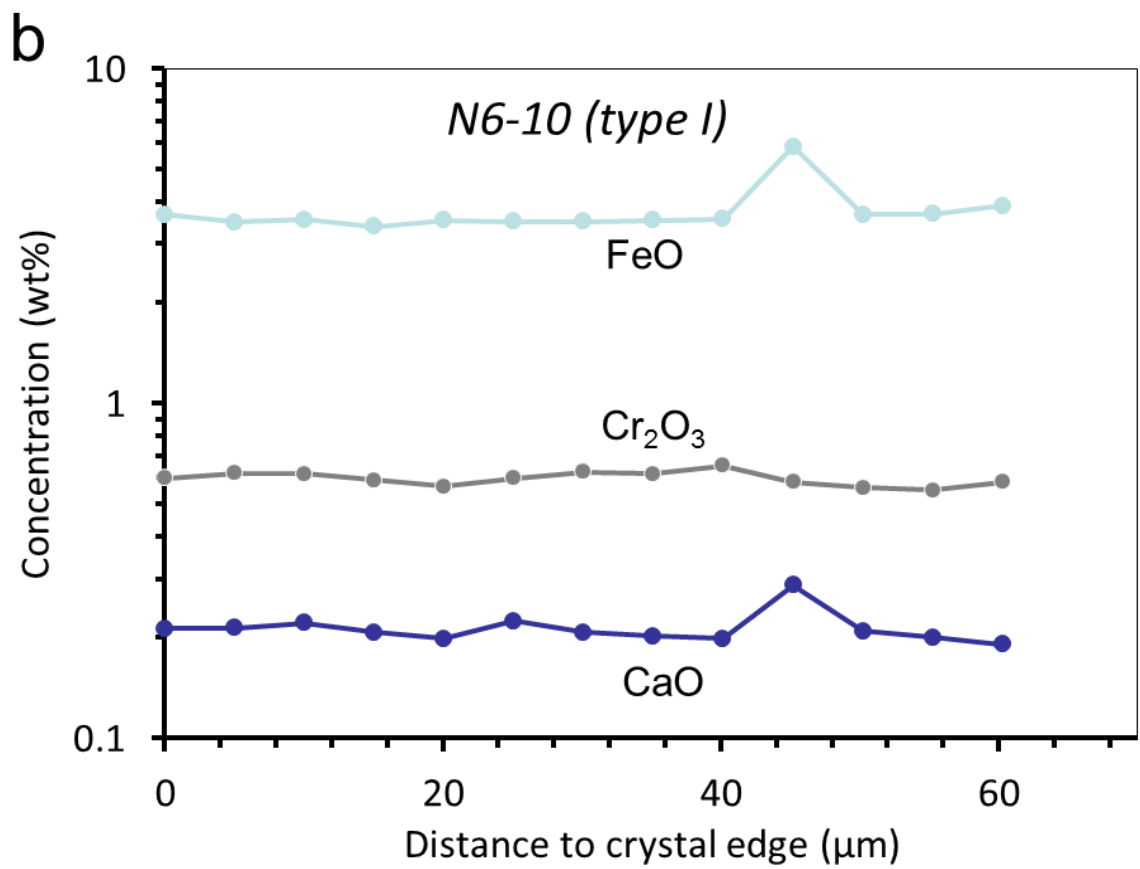

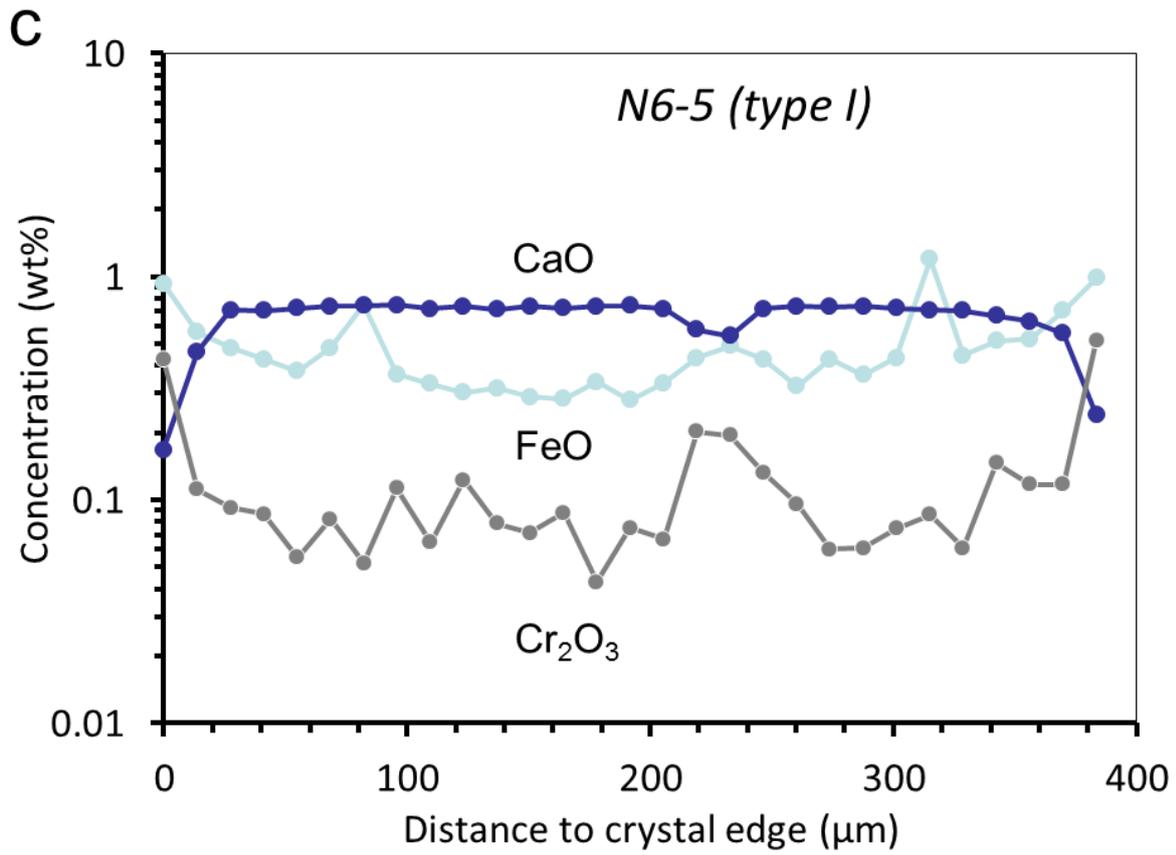
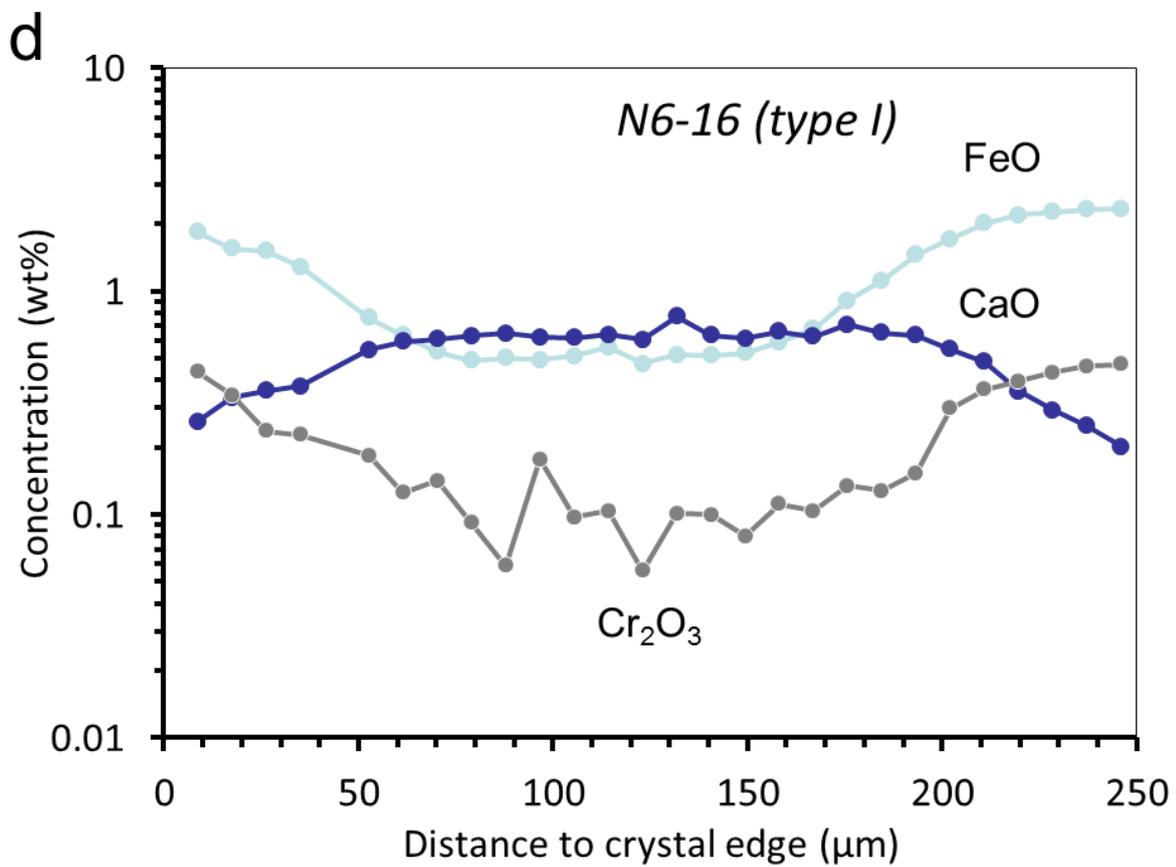

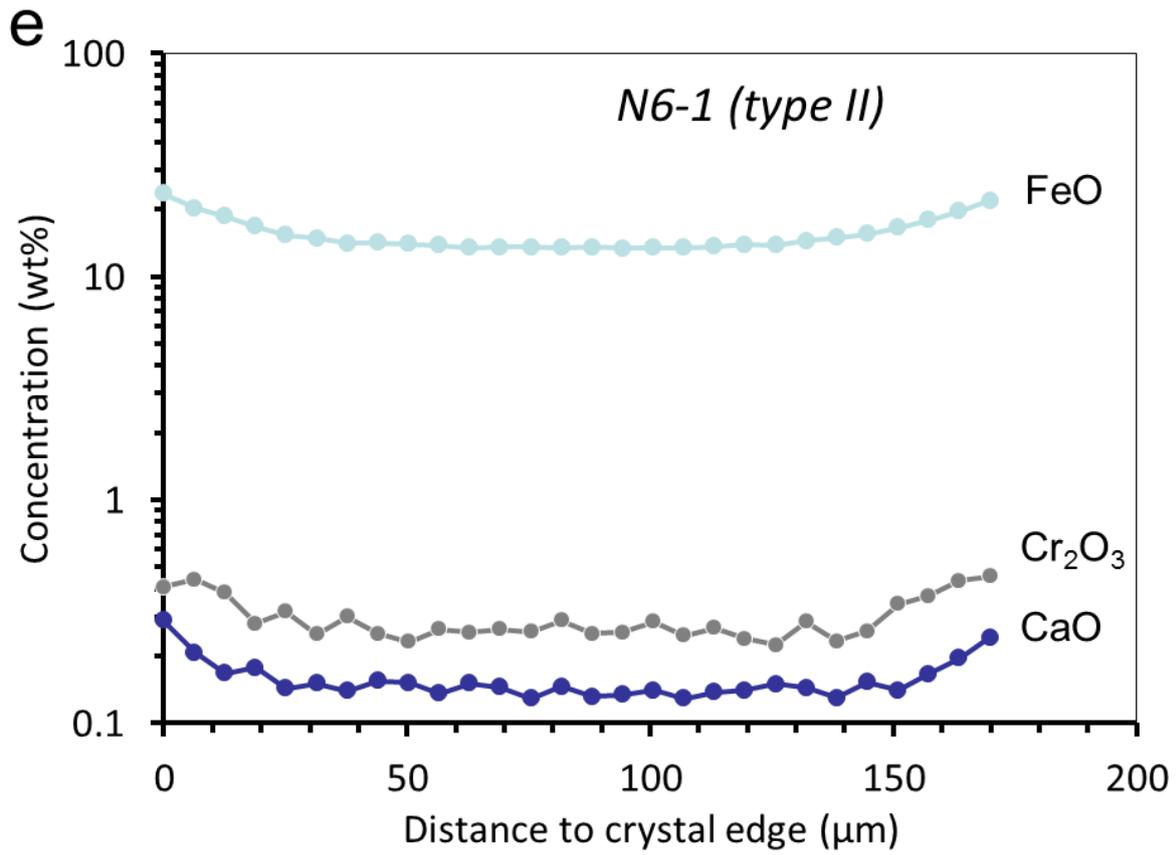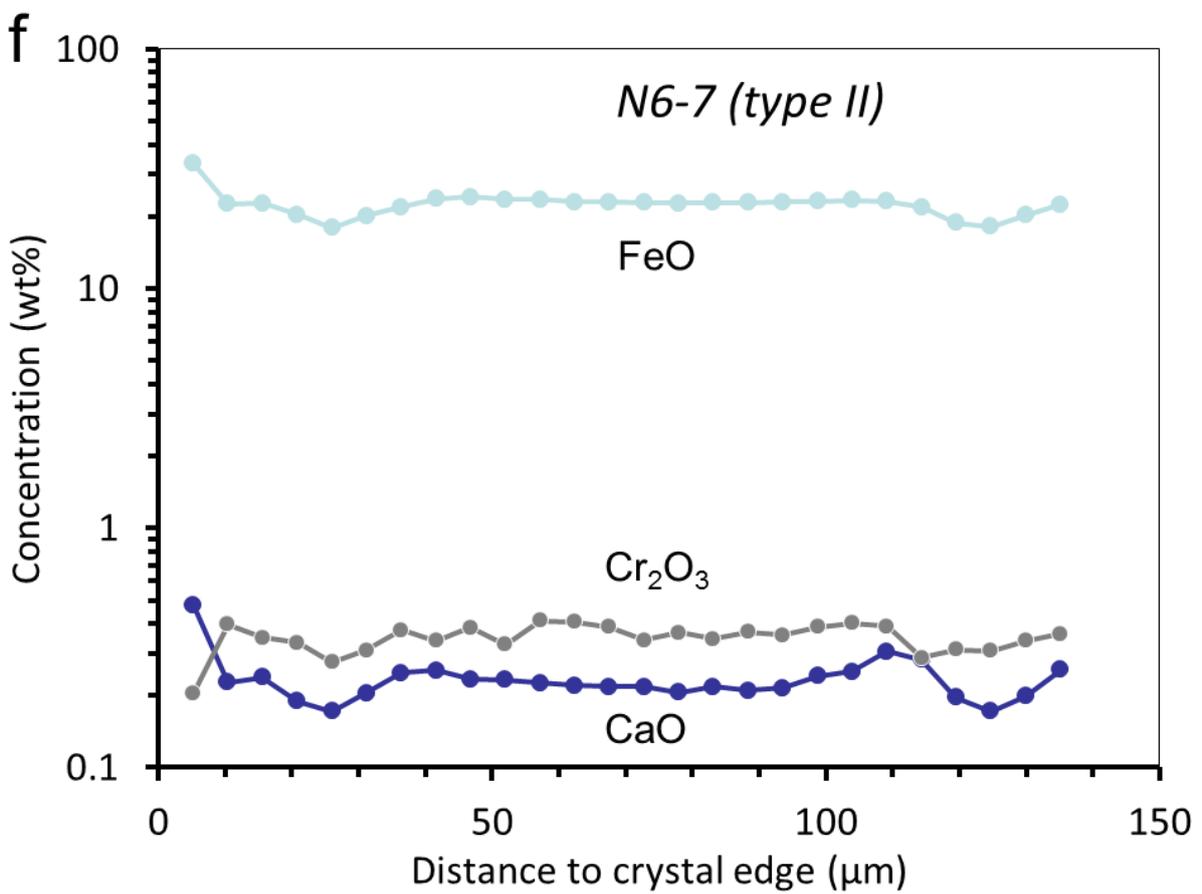

**Figure 3**: Electron microprobe traverses across selected olivine crystals in NWA 5958 chondrules. *Type I chondrules*: (a) N5-1 (Fig. 1b). (b) N6-10 (Fig. 2a of Jacquet et al. (2016)). (c) N6-5 (Fig. 1d) (d) N6-16. *Type II chondrules*: (e) N6-1 (Fig. 2f of Jacquet et al. (2016)). (f) N6-7 (Fig. 2h of Jacquet et al. (2016)).

### Chemistry

Average compositions of the different analyzed phase types are presented in Table 1, with individual analyses being listed in the Electronic Annex (Tables EA3-EA9). They are plotted in Fig. 4 for a selection of elements arranged in order of increasing volatility and Fig. 5 for Rare Earth Elements (REE). We denote by LREE and HREE the light and heavy REE, respectively. Normalization to CI chondrites (henceforth signaled by a "N" subscript) uses the values of Lodders (2003). To quantify possible anomalies, we define $Eu^*_N = (Sm_N \times Gd_N)^{1/2}$; $Ce^*_N = (La_N \times Pr_N)^{1/2}$; $Yb^*_N = (Tm_N \times Lu_N)^{1/2}$.

Analyses of chondrule silicates are quite similar to their counterparts in Vigarano and CR chondrites (Jacquet et al. 2012). Olivine and enstatite (Fig. 5a) both show steady enrichments in HREE relative to LREE (with La = 0.008-0.12 x CI and 0.08-0.4 x CI; Lu = 0.05-1.6 x CI and 0.3-1.9 x CI respectively). LREE/HREE fractionation in olivine (($Ce/Yb)_N$ = 0.007-0.7) is most pronounced in the coarsest olivine grains (Fig. 6b) as we had reported in carbonaceous chondrites (Jacquet et al. 2012); even though we did not encounter the biggest grain sizes of our previous study, hence our somewhat higher LREE concentration average in NWA 5958. As previously noted in ordinary chondrites (Alexander (1994); Jacquet et al. (2015b)), refractory incompatible element concentrations in olivine are higher, and more variable, in type I than in type II chondrules (e.g. Fig. 4, 6a). The one olivine with oscillatory zoning (apparently the first occurence in the chondrite literature despite several previous descriptions of similar zonings in pyroxene; Watanabe et al. (1986); Jones (1996); Baecker et al. (2017)), in chondrule fragment N6-7, does not exhibit a special REE pattern, with LREE being below detection. Our two augite analyses (Fig. 5c) are also fairly typical, with steady increases among LREE (La =2 and 3 x CI), pronounced negative Eu anomalies (Eu/Eu* = 0.1 and 0.2) and flattening for HREE (Lu = 19 and 4 x CI).

Chondrule mesostases show variable degrees of depletion in refractory elements (e.g. La = 3-13 x CI) and enrichment in volatile elements relative to their unaltered carbonaceous chondrite counterparts (or the pristine glass pocket in N6-1), with relatively flat, if negatively sloped patterns (($Ce/Yb)_N$ = 0.8-4). Three mesostases and groundmasses present subdued group II REE patterns with HREE depletion, negative Eu and positive Tm anomalies (Fig. 5d); similar to the mesostasis of chondrule V10 we analyzed in Vigarano (Jacquet et al. 2012)—although Tm was not measured then—, three bulk aluminum-rich Mokoia (CV3) chondrules reported by Jones and Schilk (2009) and two bulk porphyritic CM chondrite chondrules analyzed by Inoue et al. (2009). One of the chondrules in question, N1-10 (Fig. 1a), is also singled out in that its enstatite essentially mirrors the REE pattern of its mesostasis (Fig. 5d), with little inter-REE fractionation (with its EMPA-matching Al concentration of 0.7 wt% leaving little room for a possible contamination). Except for cryptocrystalline chondrule N1-3 (Fig. 2e of Jacquet et al. (2016)), bulk chondrules show LREE depletions (($Ce/Yb)_N$ = 0.3-1) and negative Eu anomalies (Eu/Eu* = 0.09-1.4) similar to bulk chondrule analyses in CM chondrites by Inoue et al. (2009).

Olivine in AOA is somewhat depleted in refractory elements and enriched in volatile elements compared to type I chondrule olivine (Fig. 4), even though no "LIME" (low-iron, manganese-enriched) olivine in the sense of Komatsu et al. (2015)—that is with Mn>Fe—was encountered. The REE pattern is also shallower (in contradistinction to forsterite in igneous FoB CAI 1623-5 in Vigarano; Davis et al. (1991)), but with a distinct depletion in LREE (($Ce/Yb)_N$ = 0.3-1.3; La = 0.08-0.2 x CI) not seen in the flatter SIMS-derived patterns of Ruzicka et al. (2012a) for CV AOAs. Otherwise, minor elements (and not too incompatible trace elements) are in broad agreement with previous data from other chondrite groups (Krot et al. (2004); Ruzicka et al. (2012a)). The spinel-diopside nodule in N5-32 and the anorthite-diopside worms in M1 show flat REE pattern (with La = 4 and 14 x CI, respectively), with the former exhibiting positive Ce and Yb anomalies (Ce/Ce* = 2; Yb/Yb* = 3.5), consistent with the mostly unfractionated bulk compositions of CV AOAs (Krot et al. 2004) as these are the main carriers of AOA REE.

Individual diopside analyses in two fine-grained CAIs (N6-9 and N6-14) show relatively flat REE patterns (La = 9 and 9.5 x CI; Lu = 4 and 14 x CI, respectively), similar to *fine-grained* CAI diopside analyses in CK chondrites by Chaumard et al. (2014)—as opposed to diopside in igneous CAIs whose patterns generally parallel that of chondrule augite (Brearley and Jones 1998)—, and consistent with the unfractionated composition of the majority of bulk fine-grained CAIs in Mighei, given the negligible REE budget of spinel (MacPherson and Davis (1994); Table EA6). The diopside spherule N6-18 shows a strong depletion in HREE (La = 4 x CI and Lu < 0.08 x CI) with positive Ce (Ce/Ce* = 2), Eu ($(Eu/Sm)_N$=1.5) and Yb (Yb = 13 x CI) anomalies, similar to the "modified group II" patterns reported in some Ningqiang (C3-UNG) CAIs by Hiyagon et al. (2011).

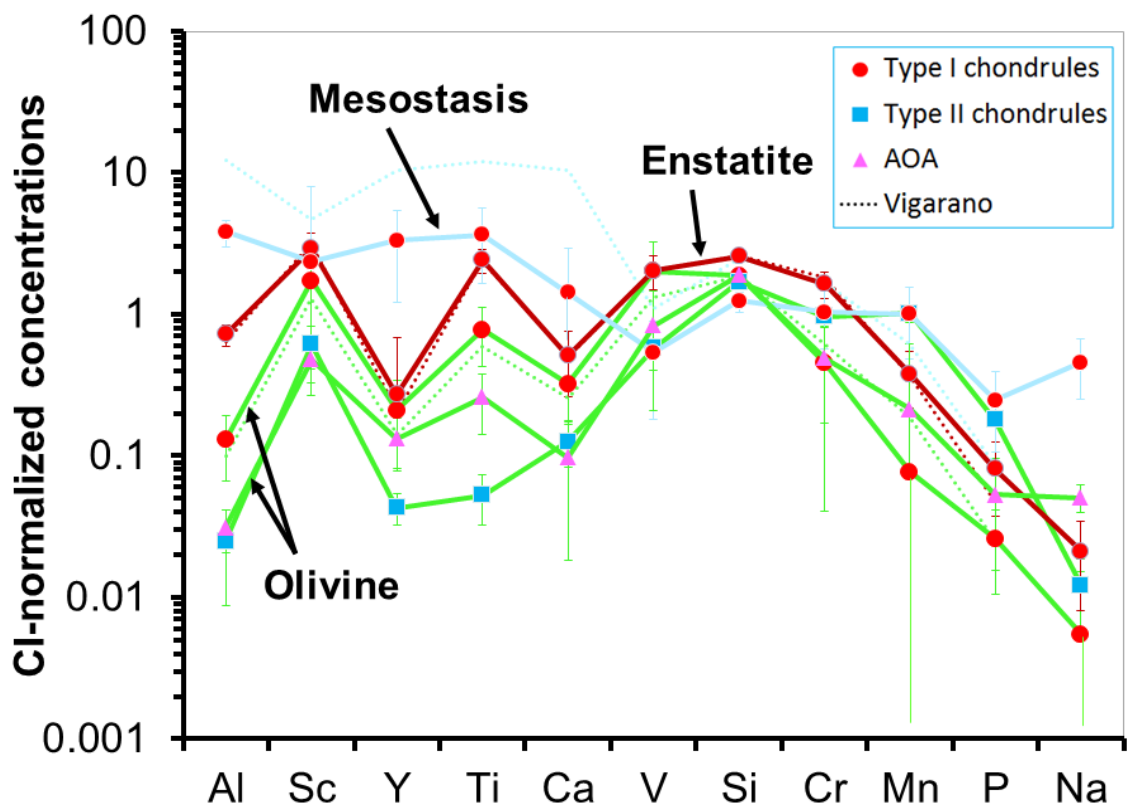

**Figure 4**: Plot of average concentrations of silicate phases in type I chondrules as well as type II chondrule and AOA olivine in an array of elements arranged in order of increasing volatility (LA-ICP-MS data except for Si (EMPA)). Chondrule types are keyed by symbol and phases by line color. Dotted lines represent type I chondrule Vigarano data (Jacquet et al. 2012) for comparison. Error bars are one standard deviation.

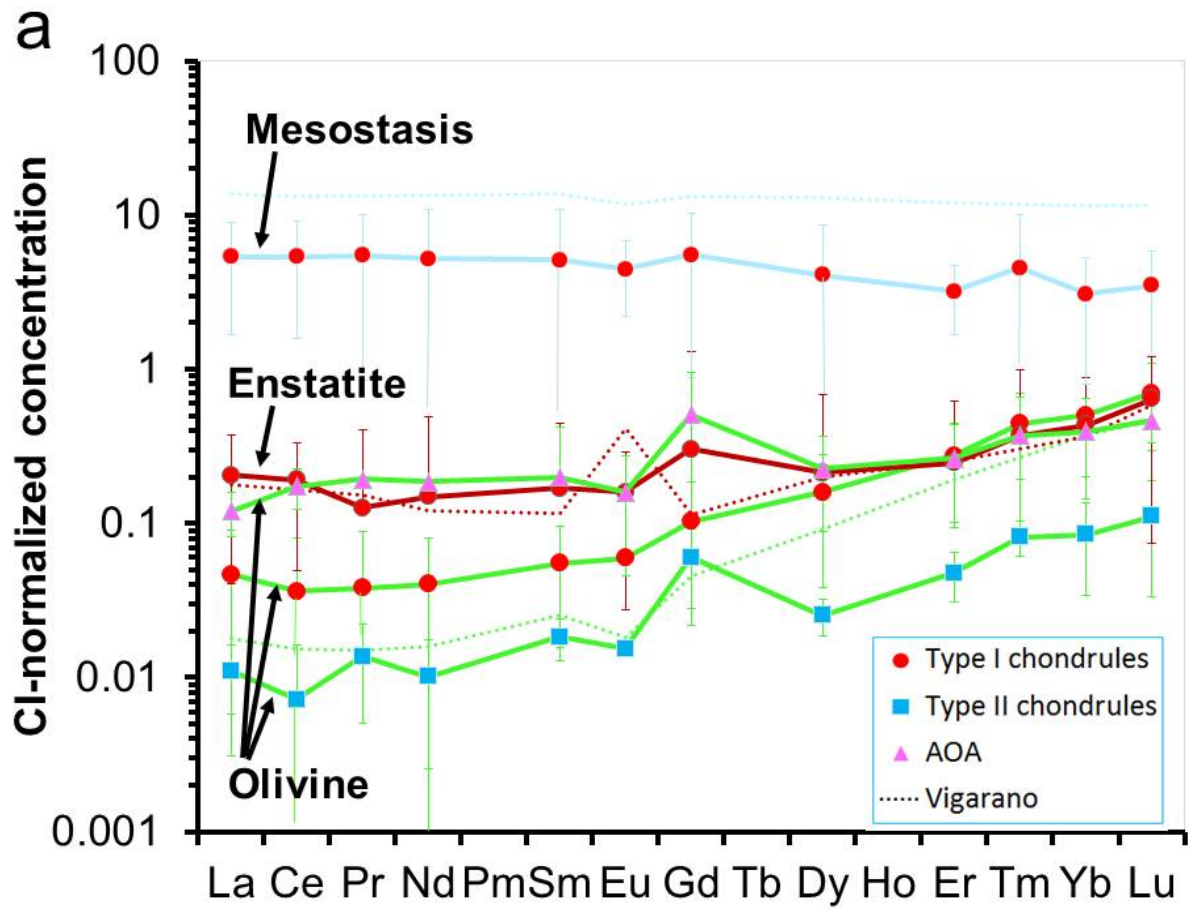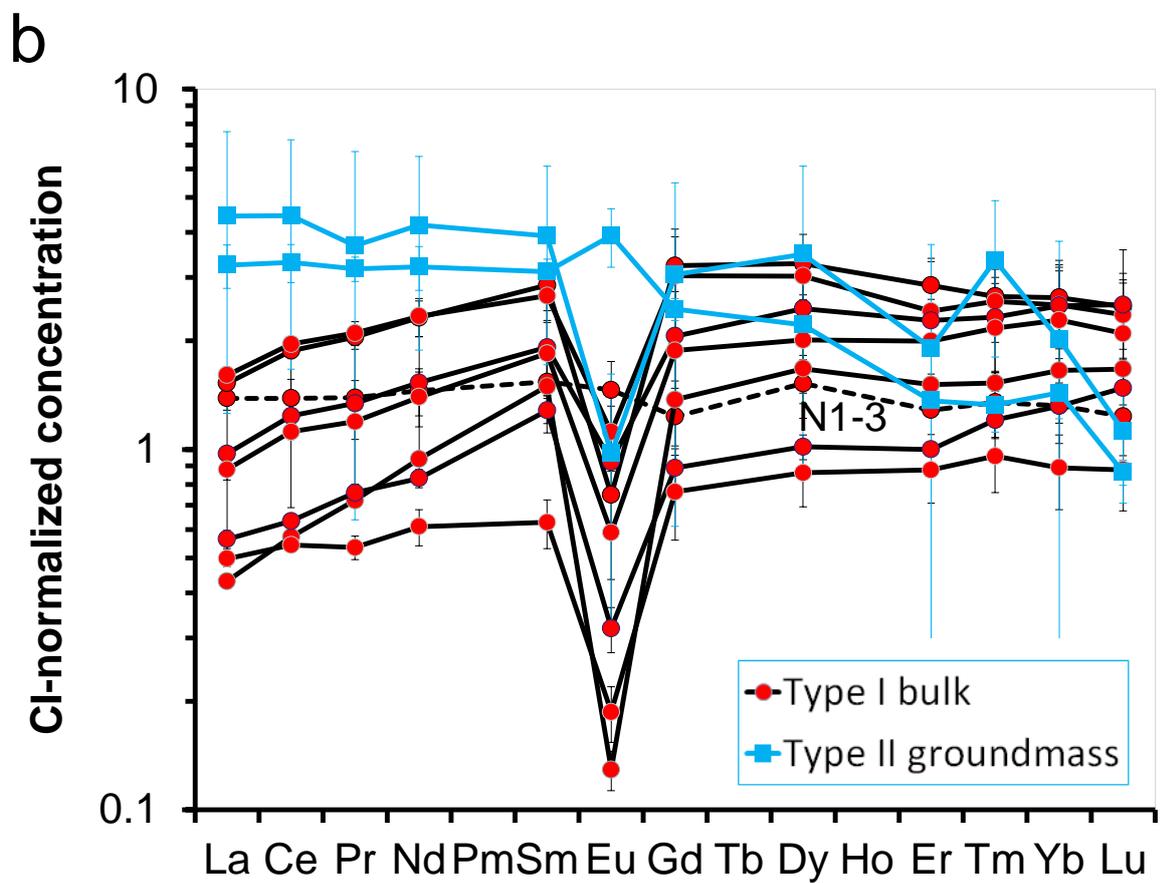

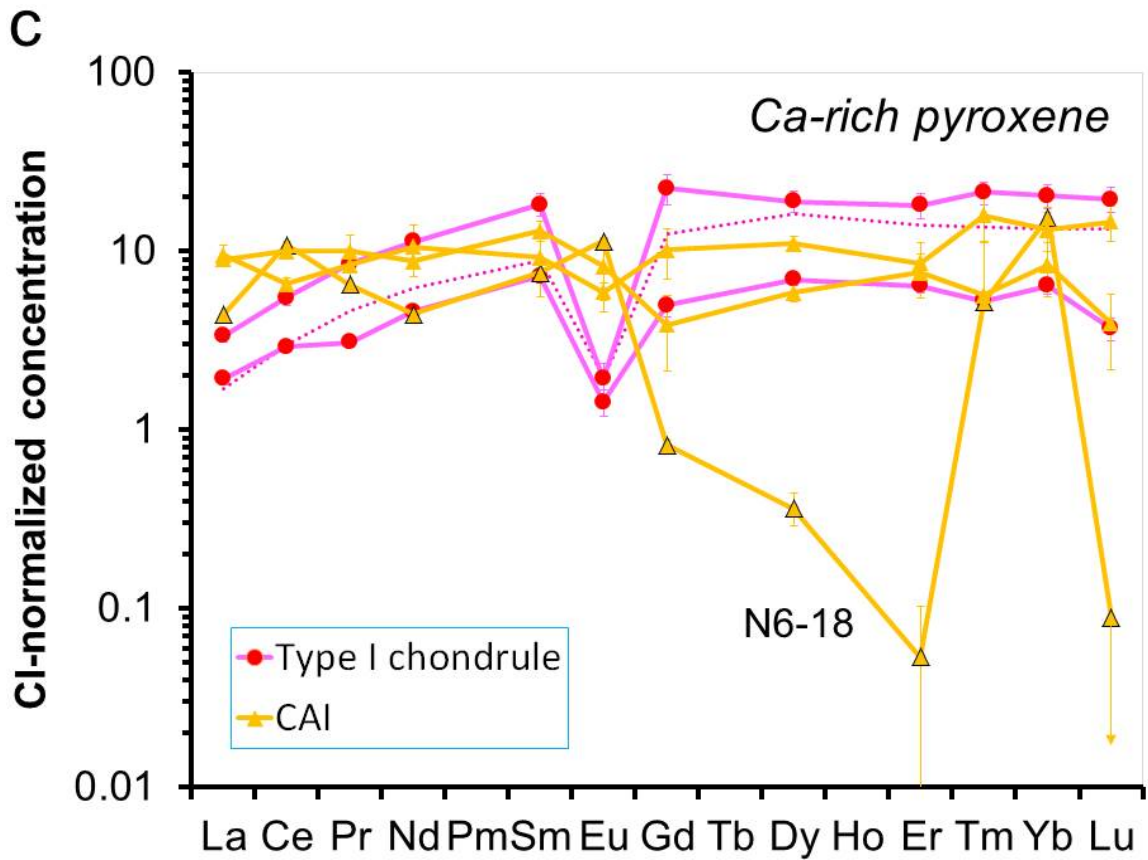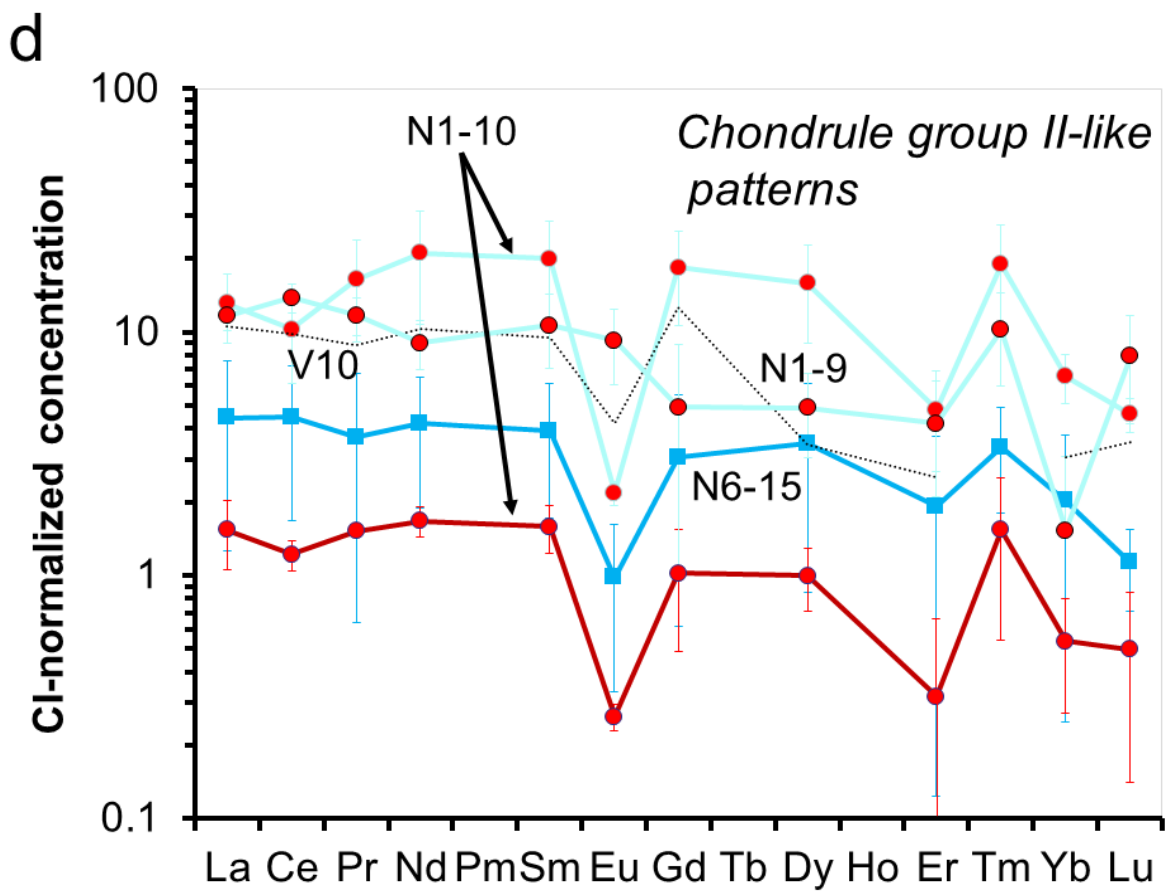

**Figure 5**: Rare Earth Element patterns. Same symbol and color coding as in Fig. 4. Dotted lines are similar data, when applicable, of Vigarano type I chondrules (Jacquet et al. 2012). (a) Average of silicate phases in type I chondrules and average olivine in type II chondrules and AOAs. (b) Chondrule bulk and groundmass analyses (dashed line corresponds to cryptocrystalline chondrule N1-3). (c) Ca-rich pyroxene in chondrules and CAIs. (d) Subdued group II-like patterns found in chondrule mesostases (light blue)/groundmasses (dark blue), with the N1-10 pyroxene pattern (brown, bottom) also shown. Dotted line is mesostasis of Vigarano chondrule V10. Error bars are one standard deviation.

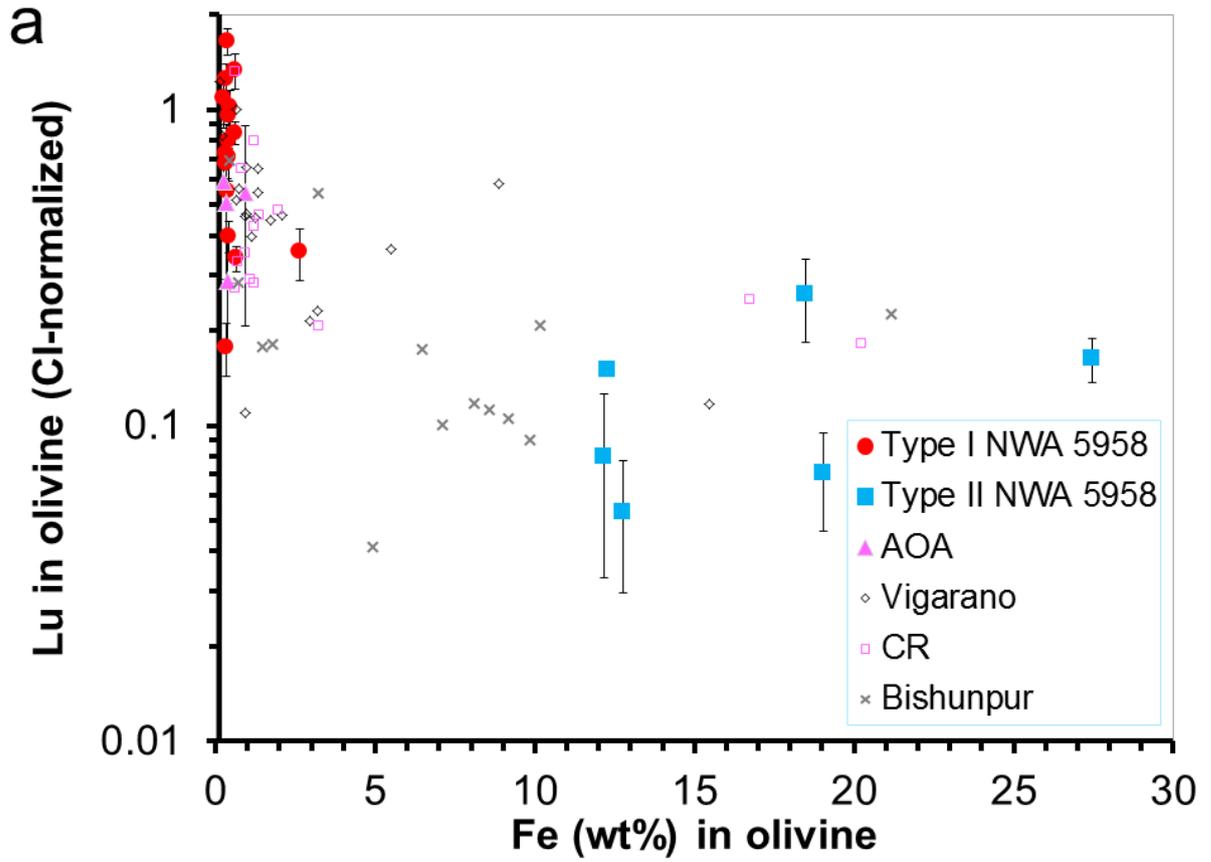

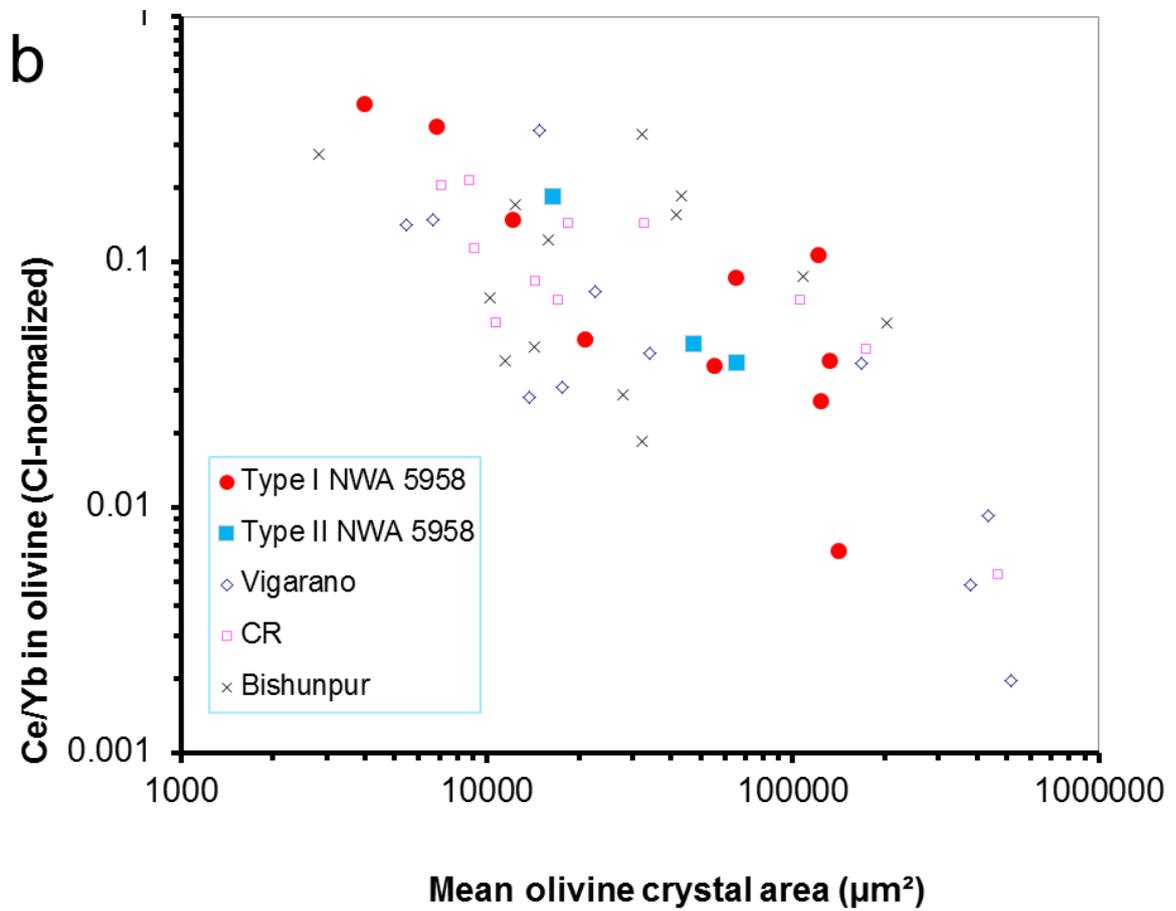

**Figure 6**: (a) Lu vs Fe in olivine. (b) Ce/Yb vs. mean crystal cross-sectional area. Also plotted are data from Vigarano, CR chondrites (Jacquet et al. 2012) and Bishunpur (Jacquet et al. 2015b). Error bars are one standard error.

Oxygen isotopes

Oxygen isotope compositions of olivine are listed in Table 2. As can be seen on the three-isotope plot (Fig. 7), chondrules and AOA scatter along the "Primitive Chondrule Mineral" line (Ushikubo et al. 2012). The $^{16}$O-rich isotope composition of AOA ($\Delta^{17}$O from -24 to -20 ‰) is similar to that of AOA in other unmetamorphosed chondrites (including CO and CM chondrites; Krot et al. (2004)). Type I chondrules have $\Delta^{17}$O ranging from -6 to -2 ‰. The plot of $\Delta^{17}$O versus Fe (Fig. 8a) is similar to data from Acfer 094 (Ushikubo et al. 2012), Yamato 81020 (CO3.0; Tenner et al. (2013)), CR3 chondrites (Tenner et al. 2015) and Murchison (CM; Chaumard et al. (2016)). These different authors noted a positive correlation among type I chondrules (with the trend flattening toward type II chondrules), but we may be witnessing in each case essentially two discrete clusters, around $\Delta^{17}$O ≈ -5 ‰ and $\Delta^{17}$O ≈ -2 ‰ (the latter being a minority among type I chondrules except in CR chondrites), respectively, with little internal correlation (the Allende type I chondrule data of Rudraswami et al. (2011), essentially confined to the former cluster, showing no clear relationship with FeO content, for example). These would correspond to the "oxygen isotope groups" of Ruzicka et al. (2007) and Ushikubo et al. (2012) as well as Libourel and Chaussidon (2011)'s modes D, E, F and A, B, C, respectively, although the latter's subdivisions of each "group" is less convincing than the hiatus between the two. Our data for NWA 5958, while broadly consistent with these literature trends, have too low statistics, given the scatter, to allow further progress on their characterization. Nonetheless, there is a clear anticorrelation between $\Delta^{17}$O and many refractory incompatible elements such as Al, Ca, REE, though not Sr, which was not apparent from EMP data accompanying the latest SIMS studies in other chondrites in the literature (Fig. 8b,c,d). Yet $^{16}$O- and CaO-rich refractory forsterites have been previously reported in carbonaceous chondrites by Weinbruch et al. (1993), Pack et al. (2005) and Ushikubo et al. (2012) and in ordinary chondrites by Ruzicka et al. (2007).

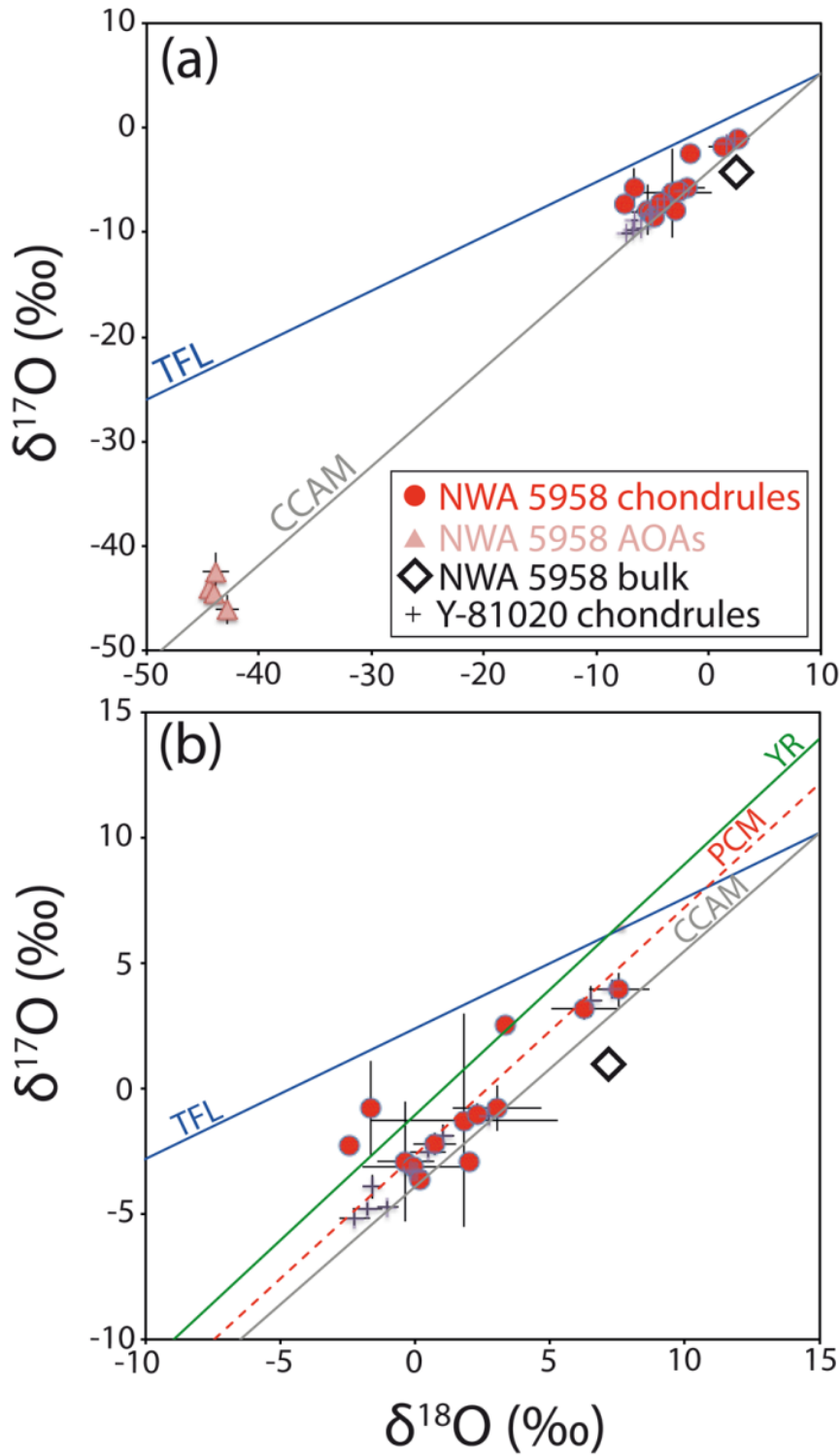

**Figure 7**: Oxygen isotope composition (with 2σ error bars) of NWA 5958 olivine (with panel (a) showing the entire range of data and (b) focusing on the chondrules), with whole-rock meteorite (Jacquet et al. 2016) and Yamato 81020 (CO3.0) chondrules (Tenner et al. 2013) also plotted for comparison. The Terrestrial Fractionation Line (TFL), the Carbonaceous Chondrite Anhydrous Minerals (CCAM) line, the Young and Russell (1998) line (YR) and the "Primitive Chondrule Mineral" line (PCM; Ushikubo et al. (2012)) are also drawn.

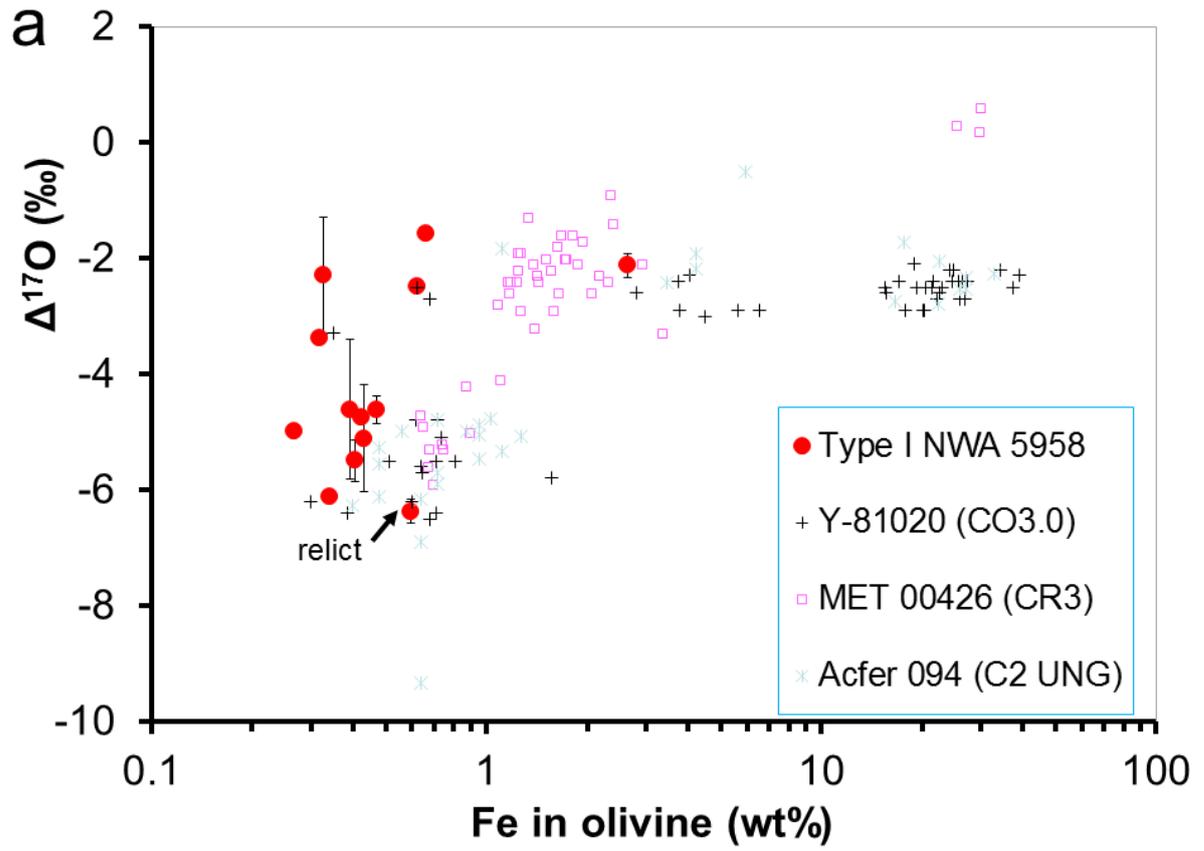
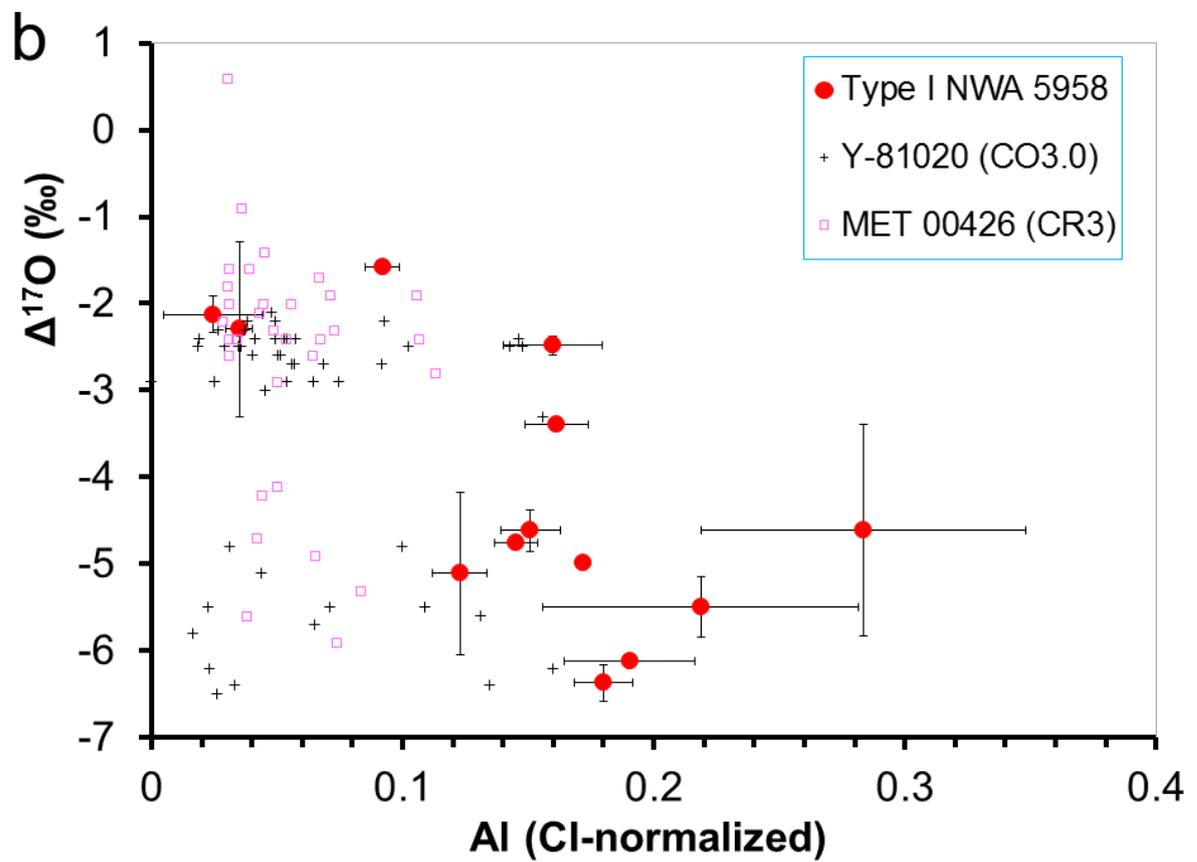

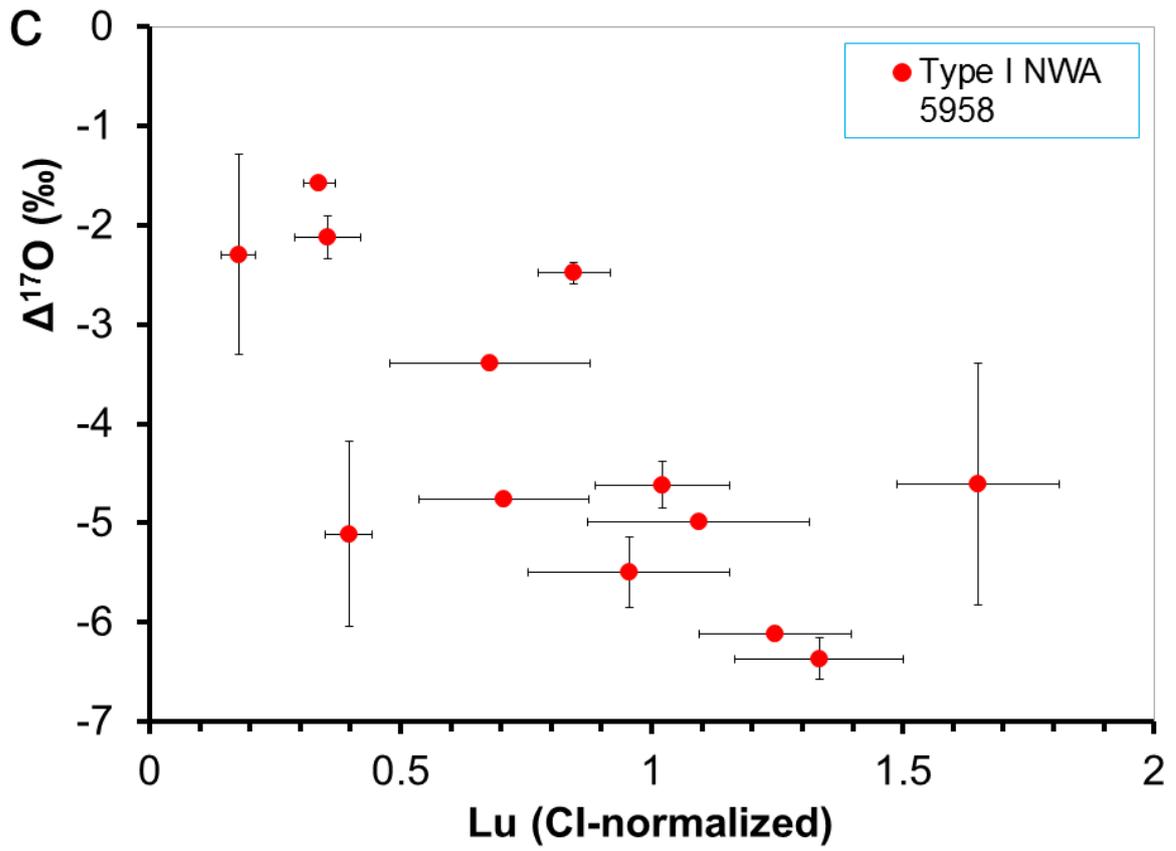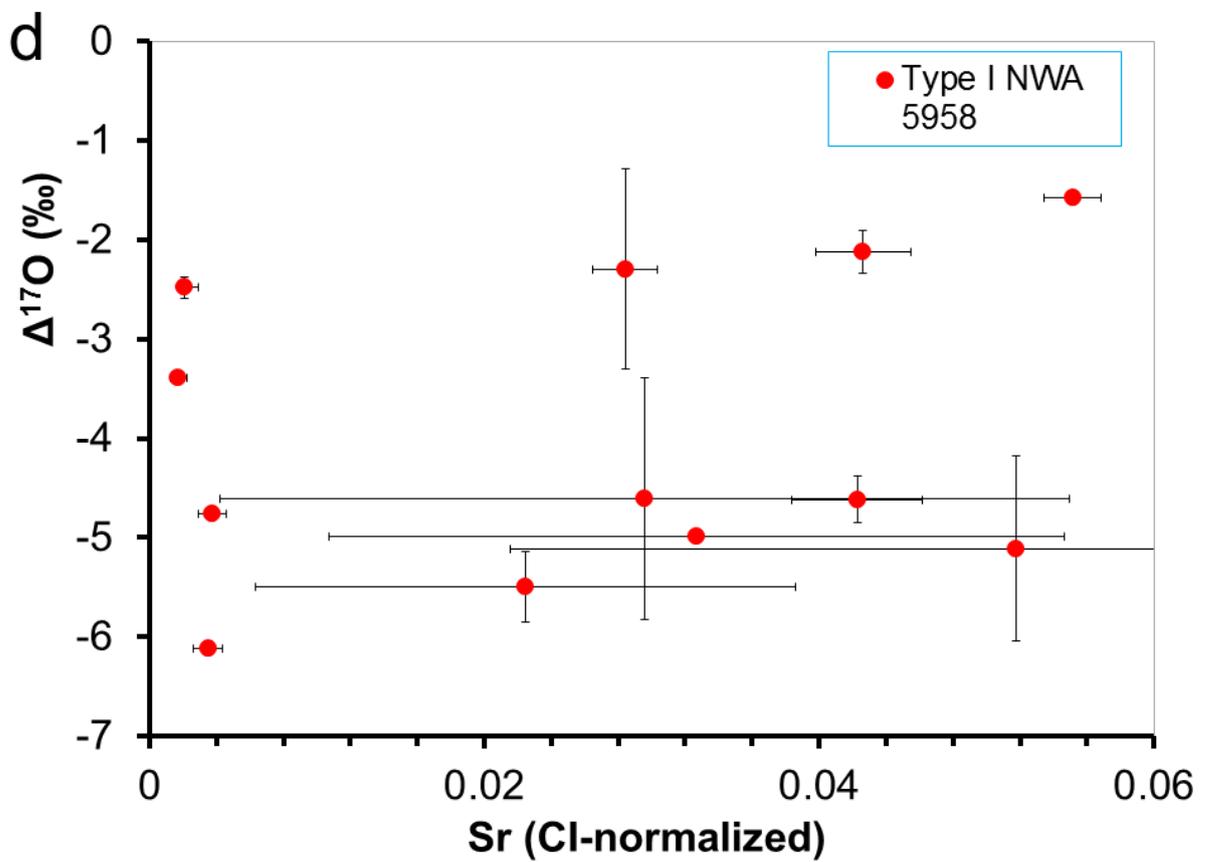

**Figure 8**: Δ$^{17}$O versus Fe (a), Al (b), Lu (c) and Sr (d) in NWA 5958 chondrule olivine. Data (with EMP analyses) from Y-81020 (CO3.0; Tenner et al. (2013)), MET 00426 (CR3; Tenner et al. (2015)) and Acfer 094 (C2 UNG; Ushikubo et al. (2012)) chondrules are also plotted for comparison when applicable. A relict forsterite (whose Fe content may have been increased by diffusion from its type II host N5-4) is pointed at in panel (a). Error bars are one standard error.

# Discussion

### Effect of aqueous alteration

Given the nonnegligible aqueous alteration experienced by NWA 5958 (Jacquet et al. 2016), it is important to evaluate how our trace element data may have been affected. Clearly the depletion of the visibly altered mesostases in many refractory elements compared to its unaltered counterparts in CV and CR chondrites (Fig. 4,5a) and enrichment in volatiles must be traced back to it. Same holds for the near-systematic LREE depletion, accompanied with negative Eu anomalies, of bulk chondrules (Fig. 5b)—unlike bulk analyses of unaltered chondrules in other chondrite groups (e.g. Misawa and Nakamura (1988); Metzler and Pack (2016))—as proposed by Inoue et al. (2009) for similar observations in CM chondrites. Yet the mesostasis itself shows little inter-REE fractionation (save for a general LREE enrichment parallel to that of unaltered mesostasis, as a result of the greater incompatibility of LREE in ferromagnesian silicates), and as such little evidence for the "tetrad effect" (that is, distinctive sub-patterns for each quarter of the REE suite as seen for some fluid-rock interactions) inferred by Inoue et al. (2009). Since (low-Ca) ferromagnesian silicates contribute negligibly to the REE budget, the fractionation seen for bulk chondrules must reflect the increasing importance, in that budget, of augite which indeed shows such patterns (see section 3.2). Indeed, augite appears to have resisted alteration much better than the now-hydrated REE-impoverished feldspathic mesostasis. As further supporting evidence, the only flat REE pattern encountered in our bulk analyses belonged to a cryptocrystalline chondrule (N1-3) devoid of augite (Fig. 5b), yet certainly no less affected by alteration than porphyritic chondrules (see Fig. 2e of Jacquet et al. 2016). While all CM nonporphyritic chondrules analysed by Inoue et al. (2009) showed flat REE patterns, barred chondrule N6-26 (Fig. 1c) does not (Fig. 5b), presumably because, as a fragment, it lacked the usual continuous olivine shell which would have protected it against alteration.

While mesostasis is clearly affected, the influence of alteration on ferromagnesian silicates seem to have been fairly minimal, judging from their overall trace element similarity with CV and CR data (Fig. 4, 5a)—consistent with the long known minor element similarity between CM and CO chondrule silicates despite the aqueous alteration of the former (e.g. McSween (1977); Desnoyers (1980); Brearley and Jones (1998)). In particular, fluid-mobile Sr averages at 0.12 and 1.92 ppm in type I chondrule olivine and enstatite, respectively, compared to 0.19 and 0.67 ppm in Vigarano. AOA olivine Sr averages at 0.61 ppm in NWA 5958, compared to 0.11 ppm in the AOA M1 in MIL 07342 (CO3.2) and the 1.18 ppm CV chondrite average of Ruzicka et al. (2012a), likewise indicating no significant depletion. The AOA M1 shows the same moderately volatile element and LREE enrichment relative to type I chondrule olivine as NWA 5958 AOA's so that these features can hardly be ascribed to metasomatism. Oxygen isotopic ratios in NWA 5958 chondrule and AOA olivine are also similar to their CO and CR counterparts; with chondrule olivine appearing even slightly *less* ferroan than in other groups for a given Δ$^{17}$O (Fig. 8), significant disturbances are thus unlikely.

## Refractory inclusions and olivine condensation

One perhaps counterintuitive property of olivine in the AOA variety of refractory inclusions is that its composition appears hardly refractory when compared to chondrules' (Fig. 4). But the paradox vanishes when it is realized that thermodynamic calculations predict the condensation of olivine at ~1300-1400 K (e.g. Komatsu et al. (2015)) while crystallization of olivine from a chondrule melt should occur over temperature intervals around ~1400-1900 K (e.g. Libourel et al. (2006); Miyamoto et al. (2009); Fedkin and Grossman (2013)). Also, dust enrichments inferred from FeO-MnO systematics in AOA olivine, about one order of magnitude above solar (Komatsu et al. 2015) are not higher than those inferred from fayalite contents for type I chondrules (e.g. Grossman et al. (2012); Tenner et al. (2013); Tenner et al. (2015)), so that moderately volatile elements would condense at lower (if anything) temperatures for AOA. The lower-temperature character of AOA olivine hence supports the condensate nature of AOAs (in contradistinction to chondrules), even though their "refractory" character resides more in their Ca-Al-rich parts (nodules and irregular patches) than in their olivine.

Another property of AOA olivine to understand is the shallow slope of their REE patterns (Fig. 5). It is sometimes tacitly assumed in the literature that near-chondritic relative ratios of some elements mark a condensation origin (e.g. Kurat et al. (1992); Weinbruch et al. (2000); Ruzicka et al. (2012a); Varela et al. (2015)), but it is theoretically not obvious why it should be so. Unfractionated abundances in a given phase *are* certainly expected if there is quantitative condensation out of a solar composition gas in the phase in question, hence, likely, the relatively flat REE patterns of diopside in the analyzed fine-grained CAIs (or the LREE portion of CAI N6-18 discussed later). This, however, would not apply to olivine where many elements, in particular REE, are much more incompatible than in diopside (or anorthite). Generally speaking, the partition coefficient D of an element between olivine and any given phase (whether a melt, gas or a mineral) can be formally expressed, along lines similar to those of section 3.11.2 of Wood and Blundy 2014), as a function of the ion emplacement strain energy $\Delta G_{strain}^{olivine}$, specifically, everything else being equal:

$$D \propto \exp\left(-\frac{\Delta G_{strain}^{olivine}}{kT}\right) = \exp\left(-\frac{4\pi E}{kT}\left(\frac{r_0}{2}(r_i - r_0)^2 + \frac{1}{3}(r_i - r_0)^3\right)\right) \qquad (1)$$

where $\Delta G_{strain}^{olivine}$ has been expressed in the framework of the lattice-strain model (Brice 1975) as a function of the Young modulus of the medium *E*, Boltzmann's constant *k*, the (strain-free) radius $r_0$ of the substitution site (~0.7 Å) and the ionic radius $r_i$ of the element of interest. Strictly, for a partitioning with diopside, the corresponding strain energy in that mineral should be subtracted in the argument of the exponential, but with a site radius ~1 Å falling squarely in the range of REE ionic radii and a Young modulus smaller than that of olivine (Wood and Blundy 2014), the correction would be subdominant.

Thus, barring any strong fractionation in the reservoir with which REE were exchanged, the dependence of the concentration in olivine on ionic radius (for a given valence) should be dominated by the right-hand-side of formula (1). So, if anything, the lower formation temperature of AOA olivine should entail a steeper dependence compared to chondrule olivine, contrary to observations, as can be seen more directly in the Onuma-like diagram in Fig. 9. While a submicron contaminant phase would be a conceivable explanation, none was seen, either inside or in between olivine crystals, in TEM by Han and Brearley (2015) in ALHA 77307 AOAs, and would anyway have little effect on Sc which prolongs the trend. One could also envision that olivine essentially acquired its REE from equilibration with the gas to circumvent the assumption of an unfractionated reservoir, but at this stage in the condensation sequence, the fractionation would be *so* pronounced (with HREE depleted relative to LREE by orders of magnitude in a non-smooth, anomaly-laden (group II-like) way; e.g. Fig. 4 in Boynton (1989)) that a much stronger departure from our prior would be expected. REE sourcing inside the AOA, in particular from diopside, thus appears more likely; perhaps the enrichment in Ca around the diopside-rich interior of AOA N1-14 may be traced to diffusion therefrom (although it may also signal a fractional condensation trend). It would seem that somehow, the formation mechanism of AOA olivine (condensation) led to anomalously less stiff (disequilibrium) partitioning behavior than usual. Anomalous partitioning behaviors were reported by Jacquet et al. (2015a) for olivine and pyroxene in the EH3 chondrite Sahara 97096, among others in LREE-enriched pyroxene spherules (a pattern also reported by Metzler and Pack (2016) in three pyroxene-rich chondrules in ordinary chondrites), which could be microkrystites, but they remain ill-understood even though incorporation of elements upon formation of a crystal was speculated to proceed differently than subsequent substitution. Thus, lack of fractionation upon incorporation of trace elements could be, after all, a manifestation of a condensation origin, but a physical rationalization thereof remains to be found.

The relatively compact textures of AOAs suggest an annealing episode, subsequent to condensation, as often inferred (e.g. Han and Brearley 2015). It may thus be wondered whether it has had any influence on the trace element budget of olivine (as opposed to initial condensation). Han and Brearley (2015) argue for subsolidus annealing periods no longer than a few tens of hours for ALHA 77307 AOAs. Diffusion over a 10 µm lengthscales would require such a duration around 1600 K for many elements, REE included (see Jacquet et al. (2015b); Chakraborty (2010)) but the typical Arrhenian dependence of the diffusion coefficients on temperature would make the required duration a factor ~3 longer for each 100 K decrement of the temperature. At face value, diffusion was likely too slow to affect the trace element budget. Yet it is quite possible that NWA 5958 AOAs were annealed for longer durations than those of ALHA 77307 so that the original olivine REE could have been as unfractionated as suggested by the analyses by Ruzicka et al. (2012a) on less annealed AOA in CV chondrites (although a marginal contamination effect (by other AOA phases) anticorrelated with grain size cannot be ruled out given the low intrinsic REE concentrations in olivine).

We close this subsection on refractory inclusions by examining the Al-Ti-rich diopside spherule N6-18. As to its (modified) group II pattern, we see no ground to depart from the conventional wisdom that it arose from condensation out of a gas from which an ultrarefractory component had been removed (if perhaps at slightly lower temperature or different oxygen fugacity than for normal group II; Hiyagon et al. (2011)), especially given that we found one example of such an ultrarefractory inclusion in the same meteorite (to be presented in a future publication). The spherical morphology of N6-18 suggests it was melted at some point. May be its monomineralic character indicates it formerly was a fragment of a diopside crystal in a coarse-grained CAI, although such textures are rare in CM and CO meteorites (MacPherson 2014), or perhaps a CAI where spinel had completely reacted away (or linger off the plane of section). The present evidence cannot decide whether the melting event took place during the CAI forming event or as part of a later (type I) chondrule forming event; the Fe content (0.4 wt%) being similar to both CAI and type I chondrule Ca-rich pyroxene. We now turn attention to ferromagnesian chondrules.

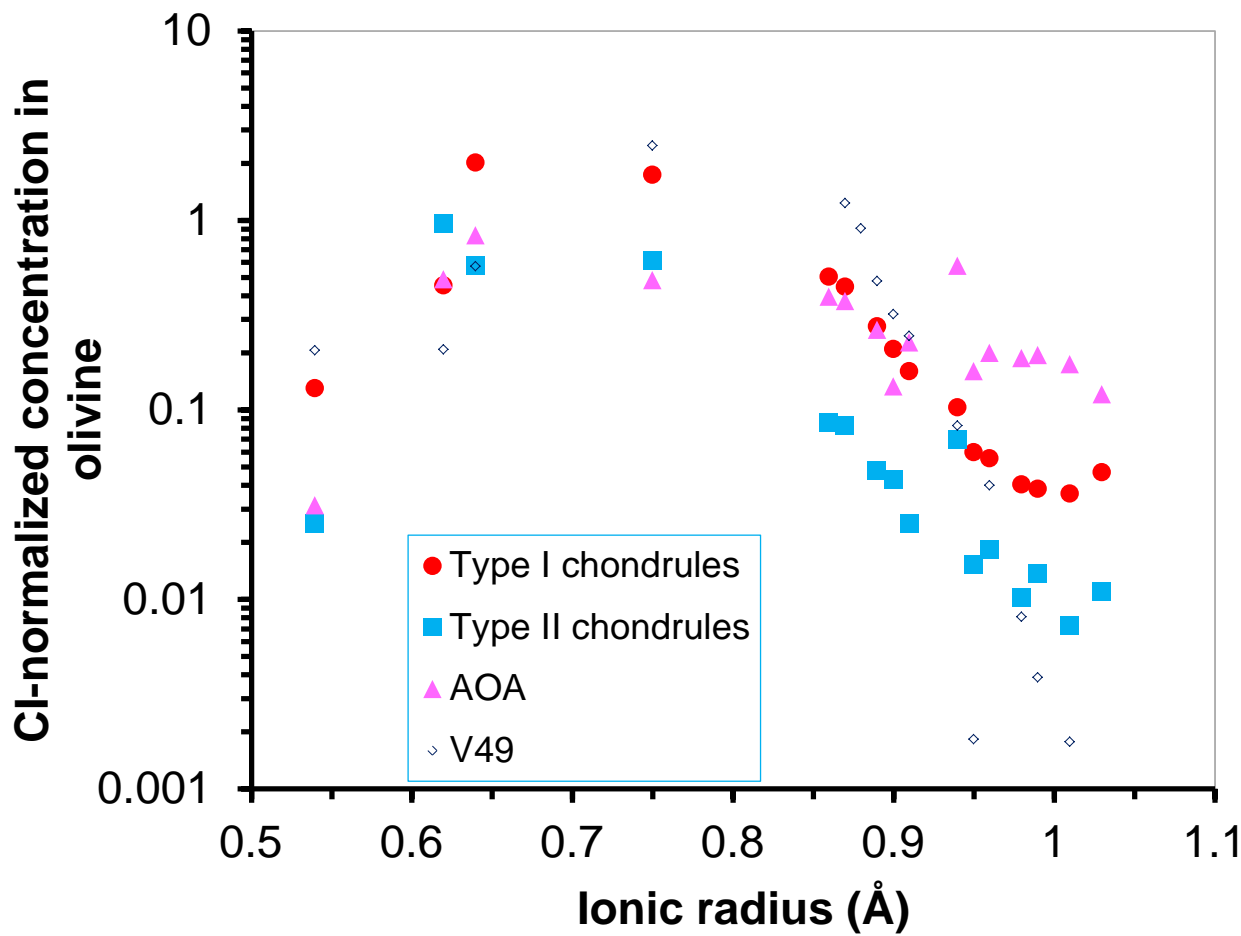

**Figure 9**: CI-normalized average concentrations of nominally trivalent elements in olivine in AOAs, type I and II chondrules, as a function of ionic radius (taken from Li (2000)). We overplot data from Vigarano chondrule V49 which best approached equilibrium igneous partitioning in our previous studies (Jacquet et al. 2012); see second part of discussion).

Chondrules and their nebular heritage

  A first remark on NWA 5958 chondrules is how unremarkable their ferromagnesian silicate trace element chemistry is compared to CV and CR chondrites (e.g. Jacquet et al. (2012)). If this is to be taken as representative of the CM-CO clan, this indicates, to first order, a geochemical uniformity of chondrule silicates, at least type I's (for distinct FeO/MnO ratios are seen in type II chondrules in CRs; see e.g. Berlin et al. 2011) in the carbonaceous chondrite superclan (disregarding, here, CH and CB chondrites). This extends to O isotopes as well, which, for type I chondrules, sample similar ranges across chemical groups (Libourel and Chaussidon 2011). Adding to this that the differences with ordinary chondrite chondrules essentially reflect the whole-rock fractionation, this is evidence for widespread common traits in chondrule-forming episodes. This does not mean that chondrules in all chondrite clans formed in a common reservoir. There certainly are petrographic differences between chondrite clans, in particular for type I chondrules (Jones 2012), suggesting that chondrule formation took place in each of their accretion reservoirs.

Despite aqueous alteration, some chondrules still unmistakably show subdued group II-like patterns (Fig. 5d). Similarly to previous authors (e.g. Jones and Schilk (2009); Inoue et al. (2009); Ebert and Bischoff (2016)), we infer that these reflect the presence of an over-average fraction of group II CAIs (perhaps of the order of $10^{-1}$) among the precursor material of these chondrules. Possibly the high mesostasis content of one of them, PP chondrule N1-10 (31 vol%; Fig. 1a), results from excess (pre-alteration) Ca, Al brought in by such a CAI; this is in line with the suggestion by Russell et al. (2000) and Krot and Keil (2002) that aluminum-rich chondrules result from over-average contributions of CAIs in their precursors (compared to mainstream ferromagnesian ones). Such interpretations are conceivable if chondrule precursors are independent stochastic admixtures of more or less refractory materials, as in a nebular setting, but not if chondrules represent large-scale averages of their native reservoir (with little fluctuations except igneous fractionation) as may be expected in many planetary scenarios (Jacquet 2014). In fact, the factor-of-a-few REE anomalies not uncommonly found in bulk chondrules (such as the subdued group II-like ones) exceed by one order of magnitude the anomalies of bulk meteorites (which average many inclusions), typically below 10 %, save for nugget or alteration effects (e.g. Evensen et al. (1978); Boynton (1984); Dauphas and Pourmand (2015)). If the latter can be taken as representative of planetesimal compositions, this is evidence against a planetary origin of chondrules, or at least a significant subset of them. While Libourel and Krot (2007) suggested a planetary origin for granoblastic olivine aggregates—themselves plausible precursors of type I chondrules (Whattam and Hewins 2009)—, it is noteworthy that one of those we analyzed previously in Vigarano (Jacquet et al. 2012), V10, also exhibited a subdued group II-like pattern (see Fig. 4d). Also, the CaO and $Al_2O_3$ contents of type I chondrule olivine (on average 0.3 and 0.2 wt% respectively) are generally much higher than in putative mantle olivine such as olivine clasts in howardites (Lunning et al. (2015); Gregory et al. (2017)) and mesosiderites (Greenwood et al. 2015) or phenocrysts in pallasites and primitive achondrites (Mittlefehldt et al. 1998), where they rarely exceed 0.1 wt%—presumably as a result of efficient extraction of basaltic melts from their crystallization regions. Whatever the formation environment of chondrules, nebular condensates may thus have been their immediate precursors.

We mentioned in the preceding subsection that AOA olivine was enriched in LREE compared to type I chondrule olivine, but in fact, even the latter is anomalously enriched in LREE compared to equilibrium partitioning expectations (Fig. 9; Alexander (1994); Ruzicka et al. (2008); Jacquet et al. 2012)). Since the reasoning above strengthens the possibility that AOAs were among chondrule precursors (see also Yurimoto and Wasson (2002); Krot et al. (2004); Whattam et al. (2010)), it is tempting to ascribe this LREE excess to direct inheritance from AOAs, incompletely expelled from the crystal lattice, as proposed by Kurat et al. (1992). Jacquet et al. (2012) countered that the coarsest olivine grains would then be expected to show the least LREE/HREE fractionation because of the longer transport lengthscale to be reached, contrary to observations (Fig. 6b), but the possibly longer durations of the high-temperature episodes responsible for these textures (Jacquet et al. 2012; see next subsection) actually make this argument equivocal. Yet it would remain difficult for the inheritance scenario to explain the lower LREE content of type II chondrules compared to type I's (if the latter were molten for longer durations; see next subsection), and the *higher* LREE contents of enstatite (Fig. 4a)—especially in N1-10—, whatever the part inherited from olivine. In fact, LREE enrichment is seen in igneous partitioning experiments, apparently as kinetic effects (e.g. Kennedy et al. (1993); Bédard (2005)), even if poorly understood (Jacquet et al. 2015a). We add to the existing constraints the fact that the one olivine which shows oscillatory zoning, in type II chondrule fragment N6-7 shows no particular LREE enrichment, suggesting little role of boundary layer effect (Albarède 2002) if the latter is responsible as suggested by Jones (1996) for similarly zoned pyroxene. Also of note are TEM observations of forsterite grains (presumably chondrule fragments) in Allende and Acfer 094 matrices by Cuvillier (2014) which reveal little evidence of a significant population of submicrometer-size glass inclusions which could have carried these excesses (without being filtered out in our analyses). At any rate, it appears that over-incorporation of very incompatible elements in olivine occurred both during AOA condensation and chondrule crystallization, and it remains a matter of speculation whether their respective physical mechanisms have anything fundamental in common.

Regardless of the behavior of very incompatible element, it remains to be understood why many incompatible elements anticorrelate with $\Delta^{17}O$ in type I chondrule olivine (Fig. 8). Since we have inferred above that CAIs were among the precursors of chondrules and since many of these incompatible elements are refractory, one might attempt to ascribe this trend to varying admixtures of refractory, $^{16}O$-rich inclusions if chondrules behaved as closed systems for all these elements. We would then expect such a correlation to apply when considering the *bulk* chemical composition of chondrules, if we reconstruct the latter from measured phase modes and compositions (as the chondrules analyzed for O isotopes have not been subject to bulk LA-ICP-MS analyses). The $\Delta^{17}O$ vs. reconstructed bulk Al plot is however inconclusive for our NWA 5958 data (Fig. 10a), for, notwithstanding uncertainties associated with 2D modal reconstructions (Hezel and Kießwetter 2010), the aqueous alteration mentioned above has visibly affected the (now-subchondritic) Al concentrations, but Fig. 10a is a pretext to graphically recall that *bona fide* bulk chondrule analyses in (comparatively pristine) Mokoia by Jones and Schilk (2009) show no correlation either. The percentage of mesostasis, which is weakly variable in Ca, Al (of which it is the main carrier phase in chondrules) in unaltered chondrules (e.g. Jacquet et al. 2012), may be a better proxy for (pre-alteration) bulk refractory element abundance of our chondrules, but the plot in Fig. 10b likewise shows no correlation with $\Delta^{17}O$. One may then envision that chondrules have been subject to open system behavior; in fact, this is suggested by evidence for influx of SiO into the chondrules (Libourel et al. (2006); Marrocchi and Chaussidon (2015); see next subsection), even though addition of material was probably too limited (<15 wt% according to Friend et al. (2016)) to significantly affect bulk refractory element concentrations. Could olivine, as a liquidus phase, fossilize the state of the precursor prior to exchange with the gas? This seems little consistent with evidence for batch crystallization (see next subsection) as well as the closeness of the O isotopic compositions and mg# of olivine and (later crystallized) pyroxene (Chaussidon et al. 2008; Ushikubo et al. 2012; Tenner et al. 2017). Following Russell et al. (2000) and Jones and Schilk (2009), we thus infer that significant O isotope exchange occurred with the gas during melting, so that the measured $\Delta^{17}O$ in olivine essentially reflects the average composition of each chondrule-forming region (rather than that of the precursor). The nature of the gas the chondrules equilibrated with remains however a matter of debate. In addition to volatile species like hydrogen or carbon monoxide, there probably were moderately volatile ones, possibly evaporated from originally condensed components (Libourel et al. (2006); Marrocchi and Chaussidon (2015)). So the oxygen isotope composition of chondrule-forming regions may have varied as mixtures between those components. Support for identification of the $^{16}O$-poor component with ice/dust (Schrader et al. (2013); Tenner et al. (2015)) comes from the positive—though not universal—correlations between $\Delta^{17}O$ and FeO content of olivine (see Section 3.3), with the latter being seen as a sensor of oxygen fugacity and thence the local solid/gas ratio. Possibly part of the scatter observed between $\Delta^{17}O$ and FeO (apart from analytical uncertainties; see Fig. 8) may arise from compositional variations of the endmembers among different chondrule-forming events, as envisioned by Marrocchi and Chaussidon (2015) and Tenner et al. (2015) for the $^{16}O$-rich and -poor components, respectively, so that actually different trends overlap in the plot. In fact, a plurality of isotopic composition vs. redox trends is certainly supported by chondrules in ordinary chondrites, where most type I chondrules lie close to the terrestrial fractionation line (Ruzicka et al. (2007); Kita et al. (2010); Ruzicka (2012)).

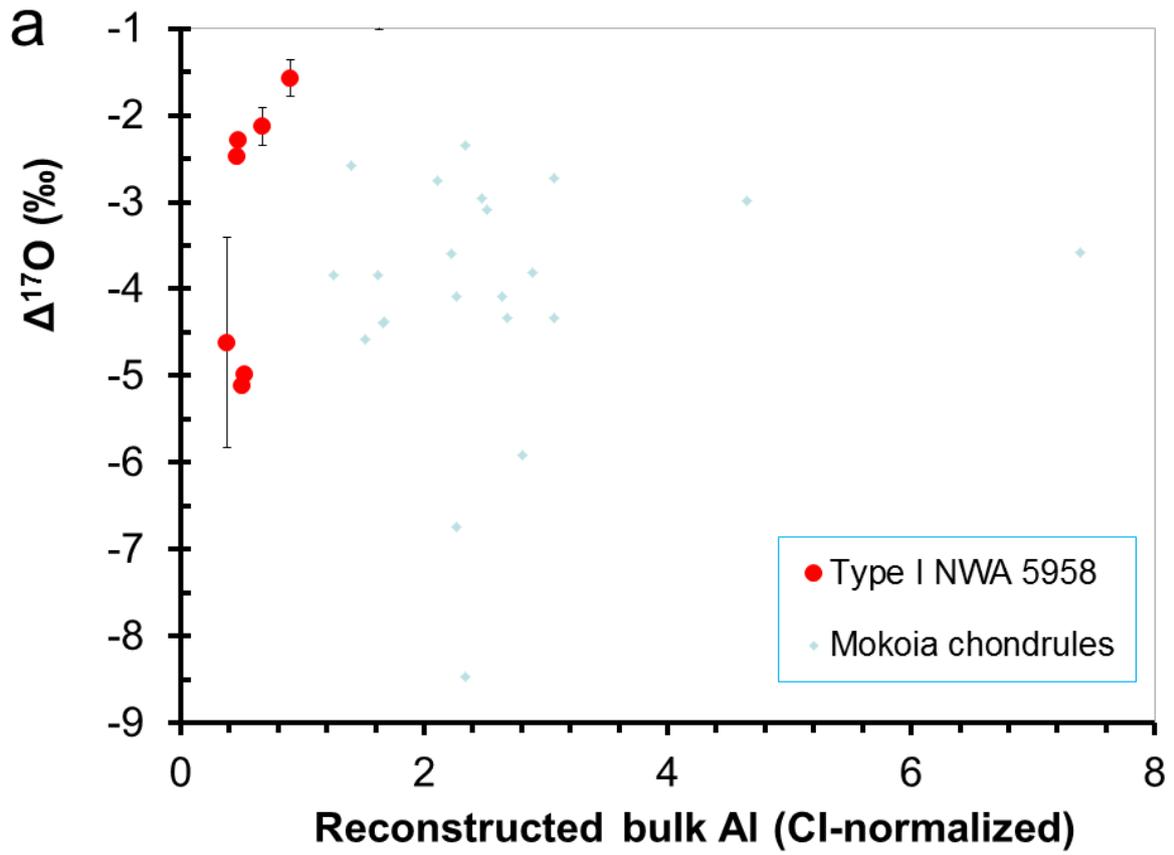

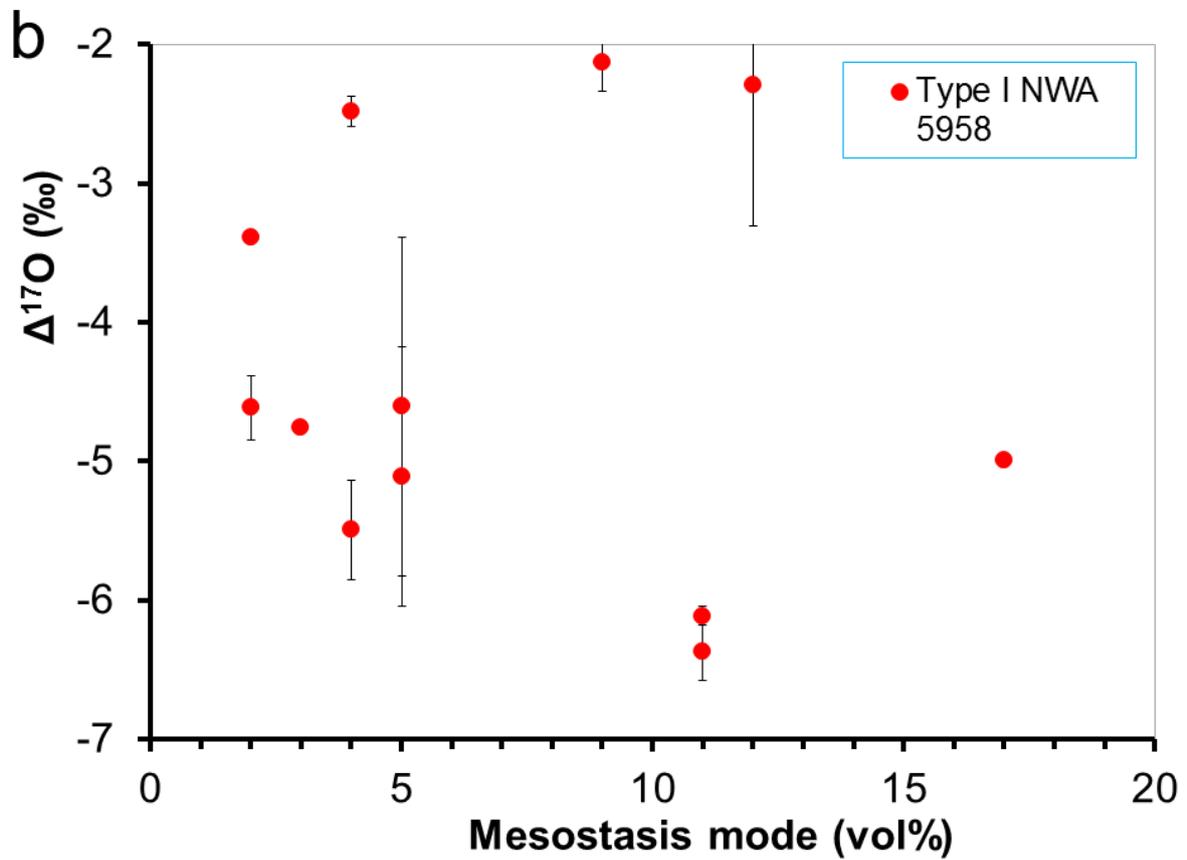

**Figure 10**: $\Delta^{17}O$ versus (a) reconstructed bulk Al and (b) mesostasis mode. In panel (a), data from bulk

chondrules analyzed in Mokoia (CV$_{ox}$3), unaffected by alteration, by Jones and Schilk (2009) are also shown. Error bars are one standard error.

### Chondrule olivine crystallization trends

Whatever the exact mechanism, if the O isotopic composition is externally buffered, it may appear paradoxical that it should at all relate to olivine geochemistry (apart from redox-sensitive or volatile elements). We thus must reflect on the determinants of trace element concentrations in olivine before trying to relate them to the isotopes. By definition, the concentration of a given element in olivine is:

$$C_{Ol} = DC_l \qquad (2)$$

where D is the olivine/melt partition coefficient and $C_l$ the concentration in the last melt with which the olivine (or more precisely, the olivine cores) equilibrated. In their study on ordinary chondrite chondrules, Jacquet et al. (2015b) agitated the question whether this melt should be identified with the initial bulk chondrule composition, as in a fractional crystallization scenario, or, rather, the mesostasis considered as the final liquid, as in a batch crystallization scenario (with intermediate solutions also possible). They noted that type I chondrule olivine had systematically higher refractory incompatible element concentrations (by a factor of ~2 on average, but up to one order of magnitude) than its type II counterpart (Fig. 3a, 4a), while little difference is seen in the average mesostasis or bulk chondrule (Gordon 2009) concentrations (although Metzler and Pack (2016) find 21-35 % depletions in refractory elements for the most FeO-rich type II chondrules in ordinary chondrites). Since little difference in crystallization temperature is inferred by Hewins and Radomsky (1990) between the two types, and, if anything, higher FeO concentrations in type II chondrules would *increase* D in the latter (Jurewicz and Watson (1988); Bédard (2005); Wood and Blundy (2014)), they concluded that the melt in equilibrium with type I chondrule olivine represented a more evolved stage in the crystallization sequence than that for (probably fractionally crystallized) type II chondrules. This would require slower cooling for the type I chondrules, perhaps over timescales of days, to ensure continued equilibration.

If we follow this line of reasoning, we have yet to understand why olivine shows more incompatible element *variability* in type I than in type II chondrules (Fig. 4a). The restricted variability of olivine in type II chondrules, if we assume that they essentially all belong to the asymptotic regime of fractional crystallization (that is, $C_{Ol}=D(T_{liquidus})C_{bulk}$), may be understood when we realize that most of the incompatible elements in question (especially refractory ones like Ca or Al) are mutually correlated at the bulk chondrule scale (e.g. Jones 2012). Then, the higher the $C_{bulk}$ for those, the higher in particular the Ca and Al content, the lower the liquidus temperature (Hewins and Radomsky 1990), and thence the lower the $D(T_{liquidus})$, which should roughly compensate the effect on $C_{Ol}$ (still, the highest FeO contents may lead to an unchecked rise of D and thus $C_{Ol}$ as can be seen in the upturn in the CaO vs FeO plots of Villeneuve et al. (2015) or Schrader et al. (2013)). But how about type I chondrules and their one order of magnitude range in concentrations? Since batch and fractional crystallizations are endmember regimes and thus allow intermediate element distributions for finite cooling rates (e.g. Miyamoto et al. 2009), one could envision that the transition between the two is recorded among type I chondrules themselves (Jacquet et al. 2015b). That is, the most incompatible-rich type I chondrule olivine (cores) would represent *bona fide* equilibrium with the final liquid while those with the lowest (core) concentrations would correspond to a significant degree of fractional crystallization. Yet there is no evidence from our EMP traverses that the latter are more zoned (or rather, less unzoned) than the other; in fact deviations from homogeneity are observed in the most incompatible-*rich* grains (and their outward drop in incompatible elements Ca and Al is little consistent with closed-system fractional crystallization of olivine). This indicates that type I olivine crystals are all equilibrated with *evolved* (as opposed to bulk chondrule composition) melts. Perhaps the evolution of these melts is tracked by the various glass inclusions, from the Si-rich ones which Faure et al. (2017) ascribed to slow-cooling entrapment of a CI (that is, bulk chondrule-like) composition melt to the Al-rich ones which Faure et al. (2012) linked to more Ca, Al-rich compositions.

But what melts? While Jacquet et al. (2015b) loosely identified these melts with the mesostases, Libourel et al. (2006) had noted that olivine in type I chondrules is out of equilibrium with its present-day composition, inconsistent with a closed-system crystallization sequence. This Libourel et al. (2006) ascribed to an influx of Si (accompanied by S; Marrocchi and Libourel (2013); Piani et al. (2016)) in the partially molten chondrules out of the ambient gas, which would have promoted the crystallization of enstatite near the margins, as commonly observed (Friend et al. 2016). The largely monoclinic structure suggests cooling rates of order 250-10 000 K/h near 1000 °C (Soulié 2014) which would also be required to avoid complete dissolution of olivine (Soulié et al. 2017). The dilution of Ca and Al in the melt following silica addition (Libourel et al. 2006) may explain the drop of Ca and Al observed at coarse olivine grain margins, which may be overgrowths following partial dissolution (Soulié et al. 2017) and/or diffusion-modified outskirts of preexisting olivine. The *increase* of Fe (and Cr) outward may perhaps point to somewhat more oxidizing conditions at this stage, as suggested by the relatively lower mg# of enstatite compared to olivine (Jacquet et al. 2016), the more ferroan composition of olivine in pyroxene-rich type I chondrules in Semarkona (Jones 1994), or the coexistence of possibly igneous magnetite in CV chondrites (Marrocchi et al. 2016).

The chemical disequilibrium between olivine and mesostasis as well as the rapid cooling rates inferred for this gas-melt interaction event (compared to the hours-days long equilibration timescales of olivine crystals by diffusion; Jacquet et al. (2015b)) indicate that olivine was then relict (Libourel et al. 2006). It is however not necessarily implied that olivine formation, while *predating* the gas-melt interaction, was an event *independent* thereof (as in the planetary debris picture of Libourel and Krot (2007) discussed in the previous subsection). For the sake of simplicity, but also to account for the O isotopic similarity between olivine and pyroxene (Weisberg et al. (1992); Tenner et al. (2015); but see Chaussidon et al. (2008) for some outliers), Jacquet et al. (2013) preferred a single event, with a protracted sub-isothermal phase where olivine would have crystallized followed by a "quenching" (minutes to hours) phase where pyroxene formed as a result of Si addition to the melt, possibly as the chondrule was leaving its native high-temperature region. This is qualitatively similar to the temperature curve inferred by Villeneuve et al. (2015) for type II chondrules.

Whatever that may be, the present-day mesostasis would not represent the melt with which olivine last equilibrated as a whole. One prediction though, would be that this melt would be most enriched in incompatible elements in the chondrules having crystallized the highest fractions of olivine. If variations in D and subsequent olivine dissolution or volume increase of the chondrule (e.g. by recondensation) can be ignored to first order, this would translate in a positive correlation between incompatible element concentrations in olivine and olivine mode. This is approximately verified in Fig. 11 although scatter, inevitable given non-representative 2D modal data or superimposed bulk chondrule compositional variability[1], does not allow firm conclusions; still Jones (1994) had previously noted low CaO contents in IAB and IB chondrules in Semarkona compared to IA chondrules (Jones and Scott 1989). A correlation is also seen with olivine grain size (Fig. 11c, d), as olivine-rich chondrules tend to be dominated by large crystals, such as may result from protracted crystal growth and merging. The contrary behavior of Sr, compared to other incompatible element (Fig. 8d), could be now understood if one realizes that the "e-folding temperature" of its partition coefficient, $(d\ln D/dT)^{-1}$, about ~100 K (see equation (1)) is much lower than for others (e.g. ~300 K for Ca; ~700 K for Lu) and cannot be ignored; in fact the cooling-driven decrease of D would have overcome the increase in $C_l$ as olivine crystallization advanced.

---

[1] It is nonetheless noteworthy that the latter would have produced the opposite trend in a fractional crystallization scenario (as olivine mode anticorrelates with the incompatible element carriers of the bulk), lending further support to the batch crystallization hypothesis.

If the above reasoning holds, the variation of incompatible elements in type I chondrules, reflecting varying degrees of olivine crystallization, may relate to the duration of the above-discussed sub-isothermal phase if olivine crystallization was interrupted by the quenching phase. There would thus be a continuous trend (of decreasing formation duration) from the most incompatible-, $^{16}$O-rich to the least incompatible-, $^{16}$O-rich type I chondrule olivine and thence to the $^{16}$O-poor, ferroan type II chondrule olivine. If the oxygen isotopic composition is indeed related to the solid/gas ratio as entertained in the previous subsection, that would mean that the correlation between solid abundance and cooling rate inferred by Jacquet et al. (2015b) also holds in the sole realm of type I chondrules. What such a correlation would physically mean remains however dependent on a specific chondrule formation scenario, as discussed by Jacquet et al. (2015b). They commented for example that the impact jetting scenario did involve such a prediction (Johnson et al. 2015)—although the compatibility of such a scenario with the variability of chondrule compositions has yet to be investigated—, as well as the shock wave models (Desch and Connolly 2002)—although this may actually hold only for large-scale shocks, with contrary behavior seen for planetary embryo bow shocks (Mann et al. 2016). It is beyond the scope of this paper to speculate more along those lines.

Before closing, we should comment that while we have reasoned above in terms of correlations across all type I chondrules, it is possible (as alluded to in the "Oxygen isotopes" subsection) to interpret the data in terms of two discrete chondrule "subtypes", or endmembers: one with refractory, generally coarse-grained forsterite at $\Delta^{17}O \approx -5$ ‰ and another with less incompatible-element rich, slightly more ferroan olivine at $\Delta^{17}O \approx -2$ ‰, that is, Ushikubo et al. (2012)'s "oxygen isotope groups". These two subtypes might merely reflect two regimes (in addition to that of type II chondrules) of a monovariant trend as discussed above, but it cannot be ruled out that they reflect qualitatively distinct formation environments (as in the "planetary" hypothesis of Libourel and Chaussidon (2011) who invoke several unrelated parent bodies; although we argued against this particular interpretation in the preceding subsection).

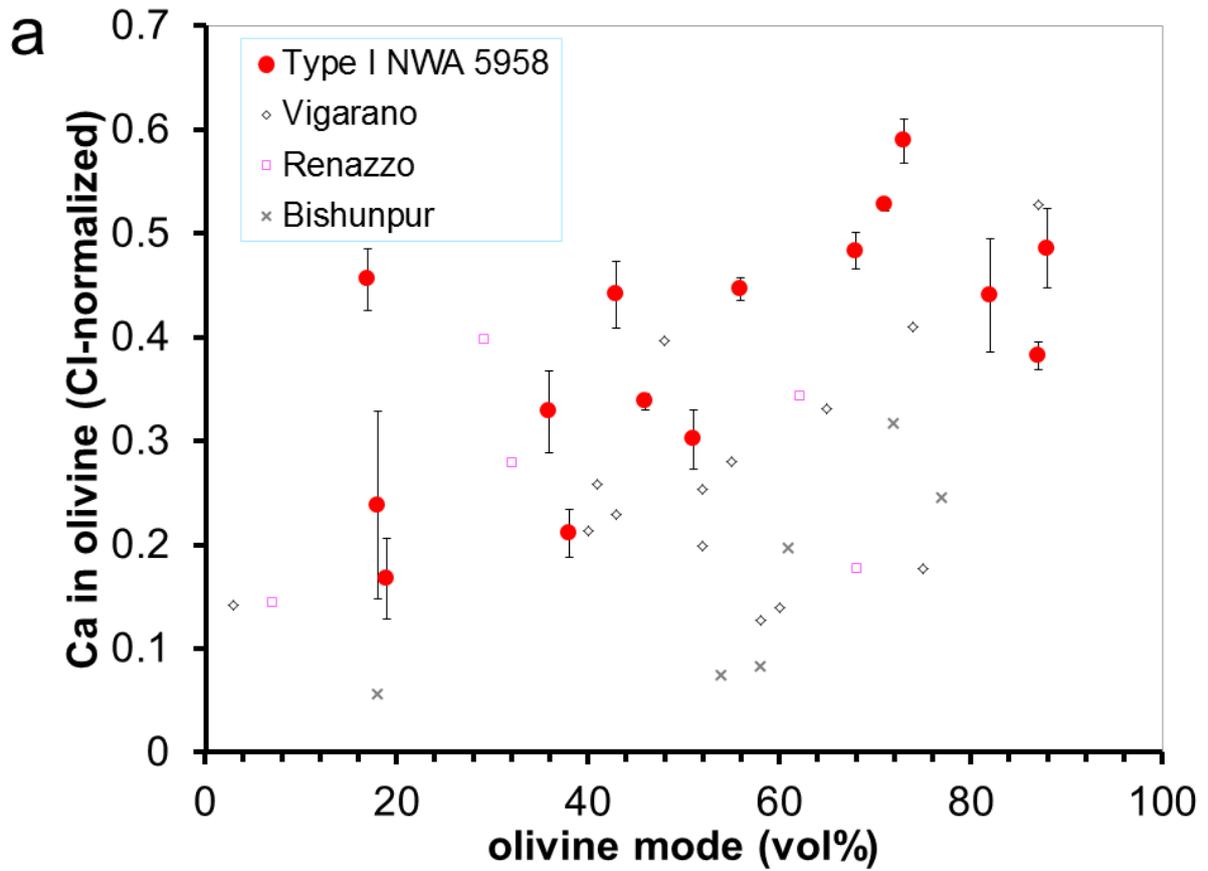

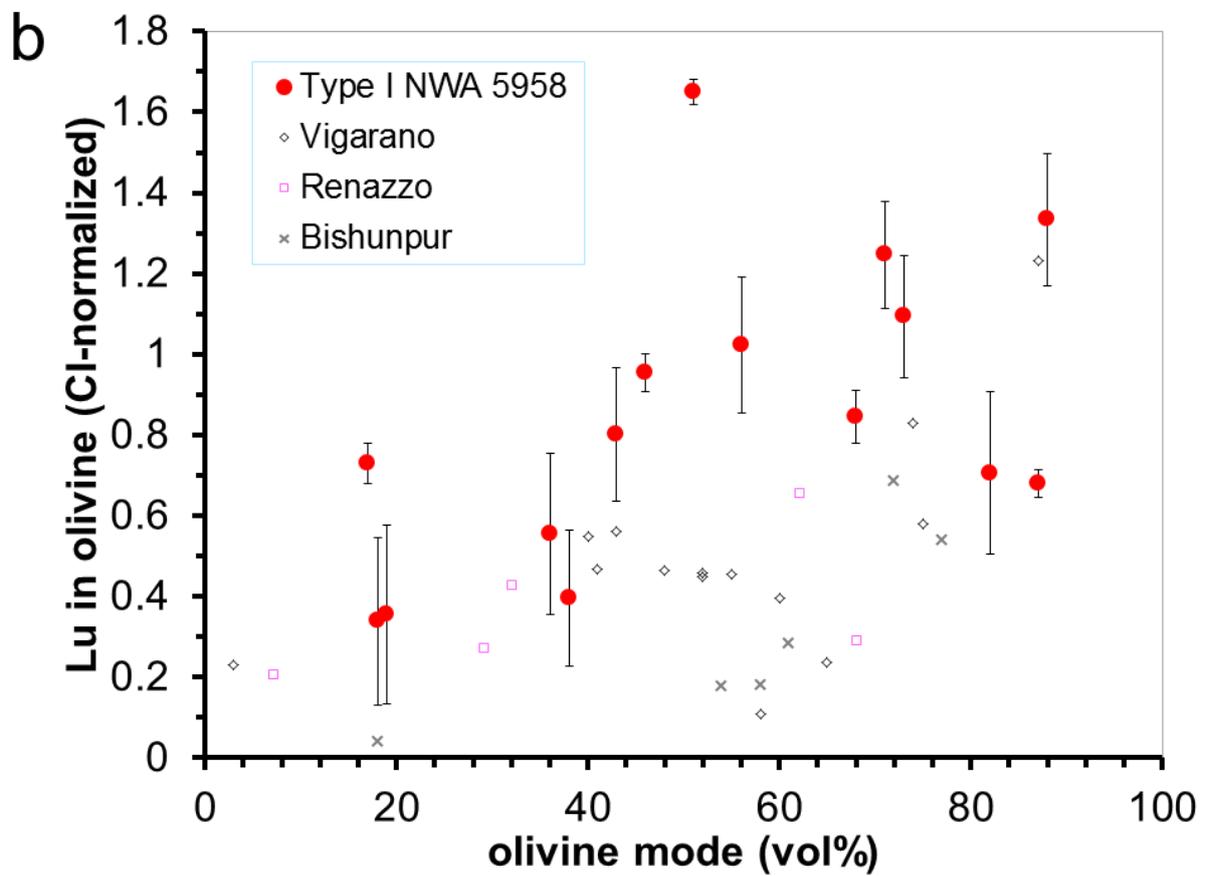

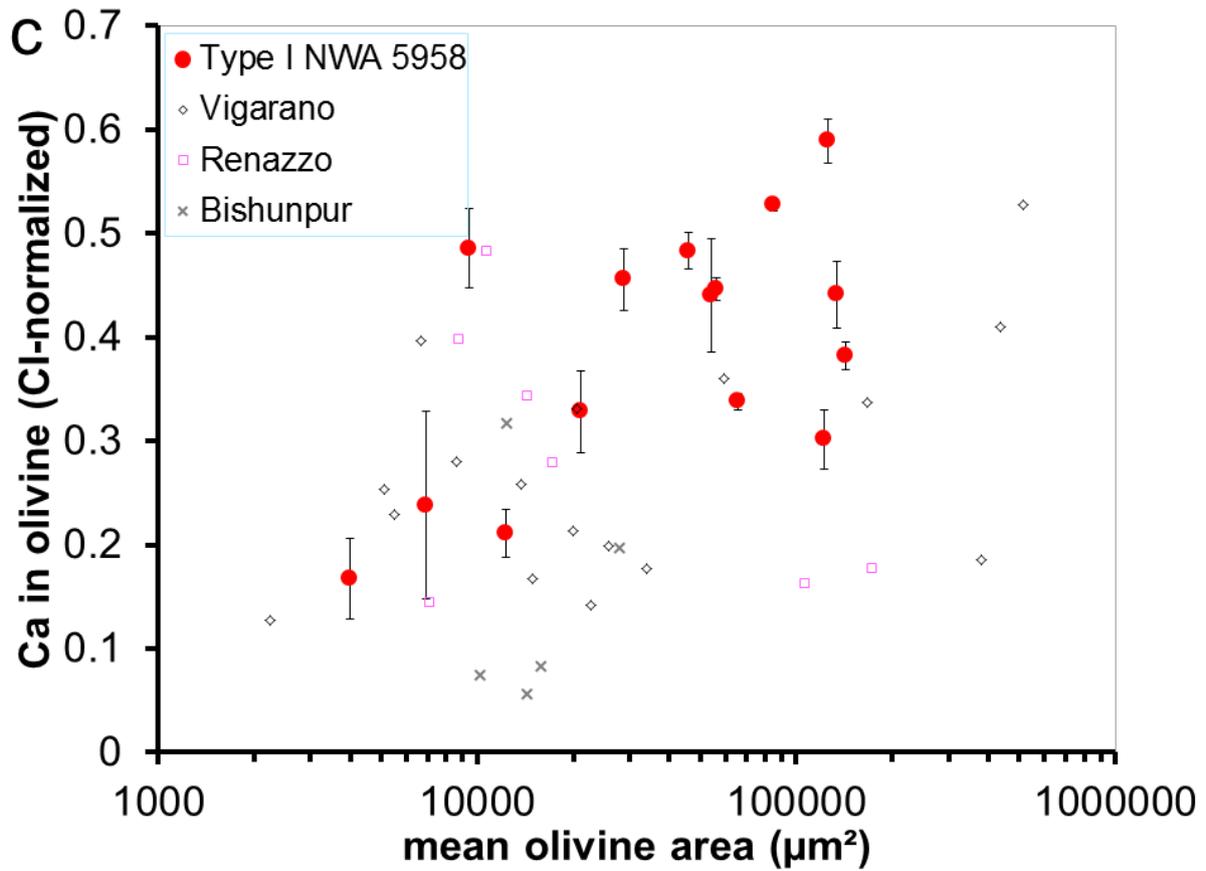
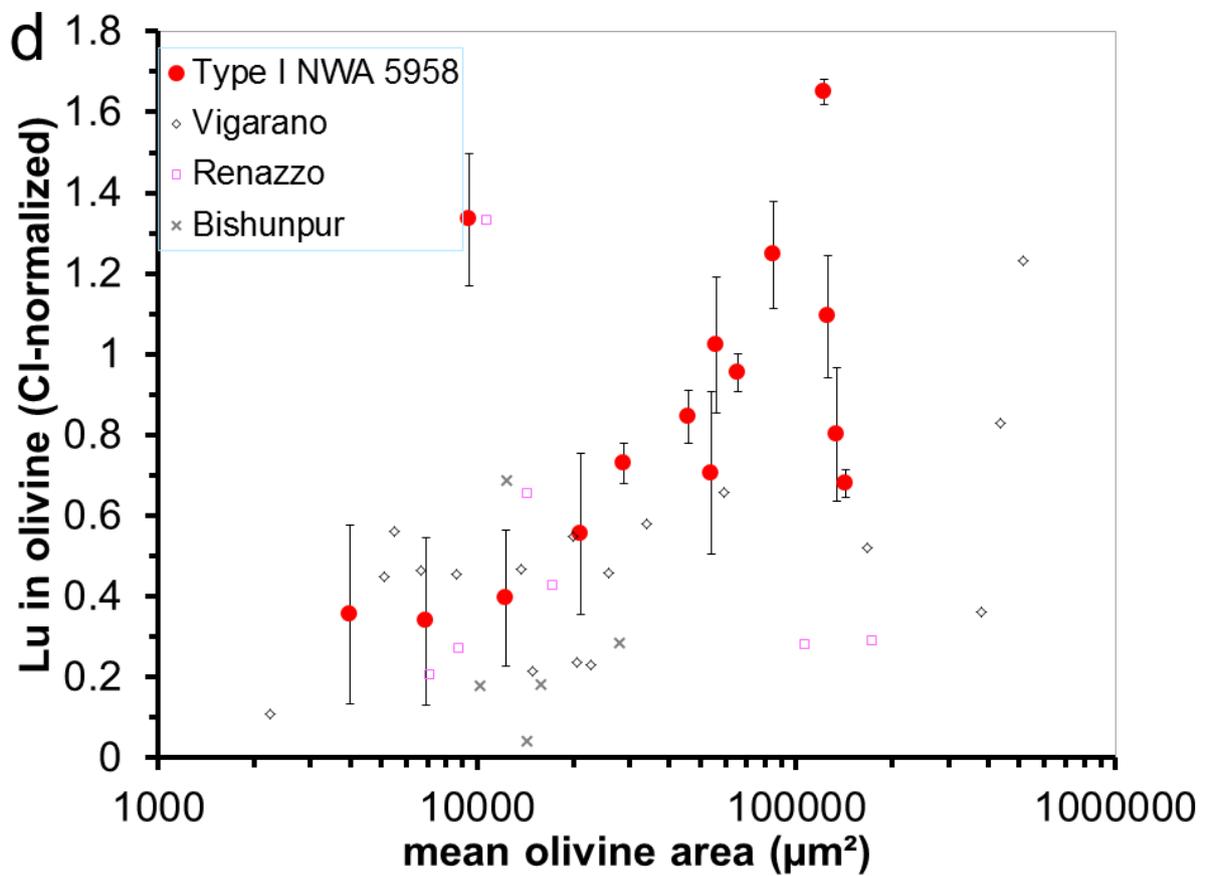

**Figure 11**: Ca and Lu in olivine versus olivine mode (a, b) and mean 2D area of analyzed crystals (c, d). Also plotted are data from Vigarano, CR chondrites (Jacquet et al. 2012) and Bishunpur (Jacquet et al. 2015b). Error bars are one standard error.

# Conclusion

We have carried out combined *in situ* SIMS oxygen isotopic and LA-ICP-MS trace element analyses in refractory inclusions (mostly AOA) and chondrules (mostly of type I) in primitive chondrite NWA 5958. Our main results are:

1. The olivine of AOA, which is as $^{16}$O-rich as their counterparts in other chondrites, is less refractory and shows REE patterns shallower than those in type I chondrules.
2. Ferromagnesian silicates in chondrules are similar to those in other carbonaceous chondrite clans, with type I chondrule olivine showing higher, and more variable incompatible element contents than its type I counterparts.
3. $\Delta^{17}$O (ranging from -2 to -6 ‰) is anticorrelated with refractory incompatible element concentrations in type I chondrule olivine, themselves correlated with olivine modal content or grain size of the chondrules.
4. Chondrule mesostases, which are typically altered, are depleted in REE and other refractory elements compared to pristine ones in other groups, with LREE-depletion and negative Eu anomalies are common in chondrule bulks. Yet some show subdued group II-like pattern. One isolated diopside spherule shows a *bona fide* modified group II pattern.

From this, we drew the following inferences:

1. AOA olivine geochemistry supports a condensate origin, although its REE pattern challenges thermodynamic predictions.
2. The occasional subdued volatility-fractionated patterns in chondrules testify to the presence of nebular condensates (in particular, but not exclusively, CAIs and AOAs) among their precursors and are inconsistent with a derivation of the latter from planetary interiors.
3. The variation of incompatible element content in type I chondrule olivine may indicate varying amounts of olivine crystallization (with correlated enrichment of residual melt in equilibrium with it), perhaps before a rapid temperature drop where gas-melt interaction promoted pyroxene crystallization.
4. The correlated variation of $\Delta^{17}$O, likely externally buffered, may then support a correlation between solid/gas ratio in the chondrule formation regions and the duration of the chondrule-forming episodes.
5. Aqueous alteration likely leached REE out of the feldspathic chondrule mesostases, imprinting the REE pattern of more resistant augite on the chondrule bulk.

*Acknowledgments*: We are grateful to the Programme National de Planétologie for financial support, to the Muséum National d'Histoire Naturelle de Paris for the NWA 5958 samples and to the "Meteorite Working Group" at the Johnson Space Center for the loan of the MIL 07342 sample. US

Antarctic meteorite samples are recovered by the Antarctic Search for Meteorites (ANSMET) program which has been funded by NSF and NASA, and characterized and curated by the Department of Mineral Sciences of the Smithsonian Institution and Astromaterials Curation Office at NASA Johnson Space Center. We thank Dr Metzler, the A.E. Dr Ruzicka and Dr Libourel for their reviews.

| Phase | Olivine | | | | | | Low-Ca pyroxene |
|---|---|---|---|---|---|---|---|
| Object type | Type I chondrule | | Type II chondrule | | AOA | | Type I chondrule |
| | Average | 1 σ | Average | 1 σ | Average | 1 σ | Average |
| Li | 0.37 | 0.12 | 0.91 | 0.62 | 2.35 | 1.22 | 2.17 |
| Na | 28 | 48 | 61 | 33 | 255 | 57 | 106 |
| Al | 1108 | 540 | 212 | 137 | 266 | 89 | 6123 |
| P | 24 | 14 | 165 | 78 | 49 | 35 | 75 |
| Ca | 2911 | 1329 | 1142 | 382 | 887 | 721 | 4624 |
| Sc | 10.12 | 8.23 | 3.60 | 1.22 | 2.81 | 1.27 | 16.94 |
| Ti | 343 | 153 | 23 | 9 | 114 | 52 | 1064 |
| V | 112 | 68 | 32 | 10 | 46 | 35 | 113 |
| Cr | 1170 | 1066 | 2507 | 379 | 1266 | 826 | 4252 |
| Mn | 147 | 231 | 1948 | 263 | 408 | 770 | 725 |
| Co | 1.97 | 4.31 | 203.39 | 88.26 | 6.34 | 3.14 | 27.56 |
| Ni | 34 | 107 | 486 | 231 | 94 | 89 | 587 |
| Cu | 0.38 | 1.00 | 0.19 | 0.13 | 2.69 | 1.51 | 5.92 |
| Zn | 1.47 | 0.71 | 1.09 | 0.44 | 4.74 | 1.71 | 5.95 |
| Sr | 0.12 | 0.14 | 0.04 | 0.05 | 0.44 | 0.35 | 1.92 |
| Y | 0.32 | 0.20 | 0.07 | 0.02 | 0.20 | 0.08 | 0.41 |
| Zr | 0.16 | 0.32 | 0.04 | 0.02 | 0.44 | 0.17 | 0.97 |
| Nb | 0.009 | 0.010 | 0.007 | 0.003 | 0.041 | 0.032 | 0.055 |
| Ba | 0.101 | 0.147 | 0.025 | 0.016 | 0.333 | 0.147 | 1.041 |
| La | 0.011 | 0.010 | 0.003 | 0.001 | 0.028 | 0.009 | 0.064 |
| Ce | 0.022 | 0.028 | 0.004 | 0.006 | 0.108 | 0.031 | 0.155 |
| Pr | 0.004 | 0.005 | 0.001 | 0.001 | 0.018 | 0.008 | 0.018 |
| Nd | 0.018 | 0.018 | 0.005 | 0.003 | 0.086 | 0.036 | 0.102 |
| Sm | 0.008 | 0.006 | 0.003 | 0.001 | 0.029 | 0.032 | 0.035 |
| Eu | 0.003 | 0.005 | 0.001 | 0.000 | 0.009 | 0.006 | 0.009 |
| Gd | 0.020 | 0.016 | 0.014 | 0.006 | 0.114 | 0.109 | 0.081 |
| Dy | 0.038 | 0.029 | 0.006 | 0.002 | 0.054 | 0.033 | 0.066 |
| Er | 0.045 | 0.028 | 0.008 | 0.003 | 0.043 | 0.028 | 0.042 |
| Tm | 0.011 | 0.006 | 0.002 | 0.000 | 0.009 | 0.007 | 0.011 |
| Yb | 0.082 | 0.049 | 0.014 | 0.008 | 0.064 | 0.041 | 0.073 |
| Lu | 0.017 | 0.010 | 0.003 | 0.002 | 0.011 | 0.003 | 0.015 |
| Hf | 0.006 | 0.012 | 0.002 | 0.001 | <0.00201 | b.d. | 0.014 |
| Pb | 0.051 | 0.035 | 0.040 | 0.018 | 0.160 | 0.097 | 0.119 |
| Th | 0.002 | 0.003 | 0.000 | 0.000 | 0.012 | 0.004 | 0.010 |
| U | 0.001 | 0.001 | 0.001 | 0.000 | 0.004 | 0.002 | 0.006 |
| Si | 199353 | 1872 | 180226 | 13433 | 200922 | 1473 | 275655 |
| **Fe** | 4136 | 1022 | 143983 | 378458 | 4684 | 2656 | 8787 |

Table 1: Summary of compositions. All concentrations are expressed in ppm (and are drawn from LA-ICP-MS analyses except for Fe and Si known from EMPA). Averaging is geometrical (standard deviation arithmetic). N1-10 is excluded from enstatite average.

| Object type | Object | Petrography | n | $\delta^{18}O$ | 1 σ | $\delta^{17}O$ | 1 σ | $\Delta^{17}O$ |
|---|---|---|---|---|---|---|---|---|
| **Chondrules** | N1-9 | I/BO | 2 | -6.65 | 0.12 | -5.76 | 0.95 | -2.30 |
| | N1-13 | I/IOG | 2 | -7.47 | 0.07 | -7.27 | 0.02 | -3.39 |
| | N5-1 | I/POP | 2 | -1.67 | 0.02 | -2.44 | 0.02 | -1.57 |
| | N5-2 | I/POP | 3 | -3.20 | 1.74 | -6.27 | 2.12 | -4.61 |
| | N5-4 | relict in type II | 1 | -2.99 | 0.10 | -7.92 | 0.16 | -6.37 |
| | N5-5 | I/POP | 3 | -5.36 | 0.53 | -7.90 | 1.20 | -5.11 |
| | N5-10 | I/POP | 3 | -5.10 | 0.92 | -8.15 | 0.23 | -5.49 |
| | N5-16 | I/IOG | 2 | -1.96 | 0.82 | -5.78 | 0.46 | -4.75 |
| | N5-21 | I/POP | 3 | -2.71 | 0.15 | -6.02 | 0.21 | -4.62 |
| | N5-24 | I/IOG | 2 | -4.81 | 0.06 | -8.62 | 0.03 | -6.11 |
| | N6-5 | I/IOG | 2 | -4.28 | 0.41 | -7.21 | 0.22 | -4.99 |
| | N6-10 | I/POP | 4 | 2.57 | 0.56 | -1.06 | 0.34 | -2.39 |
| | N6-16 | I/IOG | 2 | 1.29 | 0.62 | -1.81 | 0.22 | -2.48 |
| **AOA** | N1-8 | AOA | 1 | -44.51 | 0.10 | -44.14 | 0.16 | -21.00 |
| | N1-12 | AOA | 1 | -43.93 | 0.10 | -44.62 | 0.16 | -21.77 |
| | N1-14 | AOA | 3 | -43.85 | 0.58 | -42.46 | 0.94 | -19.66 |
| | N5-32 | AOA | 1 | -42.85 | 0.53 | -46.07 | 0.71 | -23.79 |

Table 2: Oxygen isotopic composition of analyzed objects. Shown are means for each chondrule and standard deviation of the averaged data set (n analyses; dividing by √n yields the uncertainty of the mean).